\documentclass[proof]{WileyASNA-v1_MC}

\articletype{ORIGINAL ARTICLE}

\received{\today}
\revised{}
\accepted{}

\usepackage[rightcaption]{sidecap}
\usepackage{nicefrac}

\hypersetup{
    final=true,
    pageanchor=true,
    colorlinks=true,
    breaklinks=true,
    linkcolor=blue,
    citecolor=blue,
    urlcolor=blue,
    pdfpagemode=UseNone,
    pdftitle={Characterization of chromospheric activity based on Sun-as-a-star spectral and disk-resolved activity indices},
    pdfauthor={Dineva et al.},
    pdfsubject={Solar Physics},
    pdfkeywords={Sun: activity, Sun: atmosphere, Sun: chromosphere, methods: data analysis, techniques: spectroscopic, astronomical databases: miscellaneous}}

\usepackage{textcmds}    

\usepackage[modulo, switch]{lineno}

\newcommand\arcsec{\mbox{$^{\prime\prime}$}\hspace{-0.15cm}\,}

\newcommand{\ns}{\hspace*{-5pt}}

\newcommand{\ca}{Ca\,\textsc{ii}}
\newcommand{\cak}{Ca\,\textsc{ii}\,K}
\newcommand{\cah}{Ca\,\textsc{ii}\,H}
\newcommand{\cahk}{Ca\,\textsc{ii}\,H\,\&\,K}

\begin{document}

%===============================================================================
%   TITLE, AUTHORS, AND AFFILIATIONS
%===============================================================================

\title{Characterization of chromospheric activity based on Sun-as-a-star
    spectral and disk-resolved activity indices}

\author[1,2]{E.~Dineva*}
\author[3]{J.~Pearson}
\author[1]{I.~Ilyin}
\author[1]{M.~Verma}
\author[1,4]{A.~Diercke}
\author[1,2]{K.~G.~Strassmeier}
\author[1]{C.~Denker}

\authormark{E.\ Dineva \textsc{et al.}}

\address[1]{\orgname{Leibniz-Institut f\"{u}r Astrophysik Potsdam (AIP)}, 
    \orgaddress{\city{Potsdam}, 
    \country{Germany}}}

\address[2]{\orgname{Universit{\"a}t Potsdam},
    \orgdiv{Institut f{\"u}r Physik und Astronomie}, 
    \orgaddress{\city{Potsdam}, 
    \country{Germany}}}

\address[3]{\orgname{Steward Observatory, University of Arizona}, 
    \orgaddress{\city{Tucson, Arizona}, 
    \country{USA}}}

\address[4]{\orgname{National Solar Observatory (NSO)}, 
    \orgaddress{\city{Boulder, Colorado}, 
    \country{USA}}}
    
\corres{*\email{edineva@aip.de}}

%\presentaddress{Ekaterina Dineva, Leibniz-Institut f\"{u}r Astrophysik Potsdam (AIP), An der Sternwarte 16, 14482 Potsdam, Germany}

%===============================================================================
%   ABSTRACT AND KEYWORDS
%===============================================================================

\abstract[Abstract]{The strong chromospheric absorption lines \cahk\ are tightly connected to stellar surface magnetic fields. Only for the Sun, spectral activity indices can be related to evolving magnetic features on the solar disk. The Solar Disk-Integrated (SDI) telescope feeds the Potsdam Echelle Polarimetric and Spectroscopic Instrument (PEPSI) of the Large Binocular Telescope (LBT) at Mt.\ Graham International Observatory (MGIO), Arizona, U.S.A. We present high-resolution, high-fidelity spectra that were recorded on 184 \& 82 days in 2018 \& 2019 and derive the \cahk\ emission ratio, i.e., the $S$-index. In addition, we compile excess brightness and area indices based on full-disk \cak\ line-core filtergrams of the Chromospheric Telescope (ChroTel) at Observatorio del Teide, Tenerife, Spain and full-disk ultraviolet (UV) 1600~\AA\ images of the Atmospheric Imaging Assembly (AIA) on board the Solar Dynamics Observatory (SDO). Thus, Sun-as-a-star spectral indices are related to their counterparts derived from resolved images of the solar chromosphere. All indices display signatures of rotational modulation, even during the very low magnetic activity in the minimum of Solar Cycle~24. Bringing together different types of activity indices has the potential to join disparate chromospheric datasets yielding a comprehensive description of chromospheric activity across many solar cycles.}


\jnlcitation{\cname{%
    \author{E.\ Dineva},
    \author{J.\ Pearson}, 
    \author{M.\ Verma}, and 
    \author{C.\ Denker}} (\cyear{2020}), 
\ctitle{Characterization of solar-stellar activity cycles by employing the 
    chromospheric activity S-index}, 
\cjournal{ASNA}, \cvol{}.}

\maketitle

%===============================================================================
%   INTRODUCTION
%===============================================================================

\section{Introduction}\label{SEC1}

The amplitude of cyclic variations in the solar atmosphere can be traced back to the solar dynamo mechanism. The standard model dictates that the dynamo action arises in the tachocline, which produces the large-scale global magnetic field \citep{Gilman2000, Charbonneau2020}. Once emerged at the solar surface, these fields interact with the local atmosphere and bring about the various facets of solar activity \citep{Weiss2000}. Long-term, multi-wavelength observations of activity features reveal variations consistent with the 11-year activity cycle in all layers of the solar atmosphere \citep{Ermolli2014}. In addition to annual trends, most time-series of solar activity display significant modulations because of differential rotation with respect to the formation height of the emission \citep{Bertello2012, Scargle2013}. Synoptic observations of solar activity and modelling the solar dynamo are interdependent -- like in a feedback loop, so that advances in empirical data lead to tighter boundary conditions for models and more detailed models prompt for confirmation by improved synoptic observations. 

The existence and the drivers of magnetic cycles raise fundamental questions in stellar physics \citep{Egeland2017}. Major efforts are directed to studying the relation between the observed activity and the star's intrinsic properties such as age, rotation, and spectral type \citep{Judge2012, Gondoin2018, Radick2018}. This information serves as input for models, which facilitates the inference of stellar internal structure and dynamos \citep{Baliunas1996, Pesnell2012a, Pipin2016}. The solar activity cycle is a convenient benchmark for activity variations of late-type, main-sequence stars. Although the solar cycle is well documented, questions regarding its strength \citep{Bazilevskaya2000, Schroder2017} and long-time variations \citep{Usoskin2017, Willamo2020} remain open. 

In the chromosphere, plages and enhanced active network are the most prominent \cahk\ bright features, which can induce significant changes in the visible and UV spectrum of the Sun \citep{Livingston2007, Pevtsov2014}. \citet{Babcock1955} demonstrated that bright chromospheric plage regions coincide with the locations of high magnetic field concentrations in the photosphere. The wide range covered by the formation height and the many absorption and emission features visible in the \cahk\ doublet resulted in scores of solar activity indices, which display a strong 11-year amplitude and correlate well with photospheric and coronal activity indices \citep{Livingston2007}. In addition, the relation between the \cahk\ line-core brightness and the strength of the surface magnetic field is used to reconstruct long-term magnetic variations in the absence of direct magnetic field measurements \citep{Rutten1991, Pevtsov2016, Chatzistergos2019}.   
The Mt.\ Wilson HK-monitoring project, initiated by Olin Wilson \citep{Wilson1968, Wilson1978} surveyed about 2300 stars in 1996\,--\,2002 searching for signatures of cyclic activity variation in their \cahk\ emission. A legacy of this endeavor is the $S$-index, defined as the inherent flux scale of the two instruments of the HK-project, i.e., HKP-1 and HKP-2 \citep{Vaughan1978}. This index serves as a reference standard for the strength of a star's magnetic activity, thus facilitating a direct comparison with the Sun \citep{Duncan1991, Baliunas1995, Hall1996, Schroder2018}. \citet{Sowmya2021} pursued a different approach using synthesized \cahk\ spectra to calculate a realistic solar $S$-index, which was validated with solar \cahk\ imaging and spectral data, to assess stellar magnetic activity.

Plage and active network indices are related to dispersed magnetic fields of active regions and advected magnetic elements at the boundaries of supergranular cells, respectively. The most general difference in the definition of these indices is the choice of the contrast threshold. Conversely, solar irradiance indices can also be based on magnetic field measurements \citep[e.g.][]{Chapman1986}. The \cak-plage index is the most commonly used proxy for the solar UV irradiance variation \citep{Kariyappa1996, Worden1998, Foukal2009, Singh2021b}, and it is also used to reconstruct Total Solar Irradiance (TSI) variations \citep[cf.][]{Preminger2011, Chatzistergos2020jswsc}. However, \citet{Johannesson1995} cautioned that without including the active network, the UV secular trend cannot be properly accounted for. They developed an index based on the excess emission in full-disk \cak\ images obtained at the Big Bear Solar Observatory \citep[BBSO,][]{Johannesson1998, Denker1999}. \citet{Naqvi2010} expanded this work by adding a companion index accounting for the disk-coverage of the features included in the excess brightness estimators.
 
Ground-based and space-mission instruments feed large archives of synoptic observations of the Sun. The synergy between spectral and image information, available in the case of the Sun, facilitates a global and comprehensive picture of solar activity variations. Spectrometers with high spectral and temporal resolution, e.g., the High Accuracy Radial velocity Planet Searcher \citep[HARPS,][]{Mayor2003} and the Potsdam Echelle Polarimetric and Spectroscopic Instrument \citep[PEPSI,][]{Strassmeier2015} observe the Sun and other stars with comparable spectral resolution at optical wavelengths. A spectral resolution of ${\cal R} > 100\,000$ was hitherto the domain of solar high-resolution spectrographs and spectrometers. The U.S.\ National Solar Observatory (NSO) deployed the Synoptic Long-Term Investigation of the Sun \citep[SOLIS,][]{Keller2003} instrument suite with the aim to deliver a daily set of magnetograms, filtergrams, and Sun-as-a-star spectra for selected spectral windows. In addition, the space-mission Solar Dynamics Observatory \citep[SDO,][]{Pesnell2012} reveals in full-disk images and magnetograms details and dynamics of the solar atmosphere. Recent examples of calculating long-term activity indicators and solar activity studies are presented in the works of \citet{Maldonado2019}, \citet{Chatzistergos2020} and \citet{Bertello2020}.  

Our investigation of solar activity indices presents in Section~\ref{SEC2} the Sun-as-a-star spectrometer PEPSI/SDI along with ground-based and space-mission instruments (ChroTel and SDO/AIA) monitoring chromospheric activity. The derivation of activity indices is described in Section~\ref{SEC3} for Sun-as-a-star spectra and in Section~\ref{SEC4} for chromospheric full-disk images. The two types of indices are compared in Section~\ref{SEC43}. Section~\ref{SEC5} provides an outlook of future science with PEPSI/SDI, once a longer time-series of the $S$-index will be at hand, covering at least the rise to the activity maximum of Solar Cycle~25.

%===============================================================================
%   OBSERVATIONS
%===============================================================================

\section{Observations}\label{SEC2}

This study is based on high-resolution Sun-as-a-star spectra of the chromospheric \cahk\ lines, ground-based synoptic full-disk images of the chromosphere in the light originating from the core of the \cak\ line, and synoptic full-disk UV images from space. These data are the foundation to compare solar activity indices derived from unresolved Sun-as-a-star spectra with those extracted from two-dimensional filtergrams of the entire solar disk.

%-------------------------------------------------------------------------------
%   Solar Disk-Integrated telescope
%-------------------------------------------------------------------------------

\subsection{Solar Disk-Integrated telescope}\label{SEC21}

The Potsdam Echelle Polarimetric and Spectroscopic Instrument \citep[PEPSI,][]{Strassmeier2015, Strassmeier2018} is a state-of-the-art, fiber-fed spectrograph operating in a pressure- and temperature-stabilized chamber in the basement of the Large Binocular Telescope \citep[LBT,][]{Hill2006} at Mt.\ Graham International Observatory (MGIO) in Arizona, U.S.A. PEPSI delivers solar and stellar spectra in the spectral range 3830\,--\,9140~\AA. The instrument's design provides effective radial-velocity stability. The spectral resolution is determined by the choice of the fibers, which deliver the light from the telescope to PEPSI. Daytime observations of the Sun are carried out with the highest available spectral resolution of ${\cal R} \approx 250\,000$. The 11-millimeter aperture Solar Disk-Integrated \citep[SDI,][]{Strassmeier2015} telescope mimics the binocular design of the LBT and is located on a balcony of the LBT building. The SDI telescope observes the Sun as a star and feeds disk-unresolved light to a pair of 100~$\mu$m-core fibers, guiding the light to the spectrograph.  

PEPSI utilizes two extremely large-format STA1600LN CCD detectors with 10.3k $\times$ 10.3k pixels to record cross-dispersed echelle spectra at three wavelength settings for each of the two light feeds. The pixel size is 9~$\mu$m $\times$ 9~$\mu$m. Thus, the detector area is about 10~cm $\times$ 10~cm. Considering that PEPSI can record 16 orders in the red and blue simultaneously, it is possible to acquire all 92 echelle orders over the whole spectral range in just three subsequent exposures of the solar spectrum. The typical exposure time is in the range 0.3\,--\,8.0~s, which depends on wavelength settings and sky brightness. A readout time of approximately 80~s yields a cadence in the range 80\,--\,90~s. The initial reduction of the raw echelle spectra requires an efficient data processing pipeline specifically designed for the large amount of high-resolution solar data of up to 200~GB per day \citep{Strassmeier2018}. 

\begin{figure*}[t]
\includegraphics[width=\textwidth]{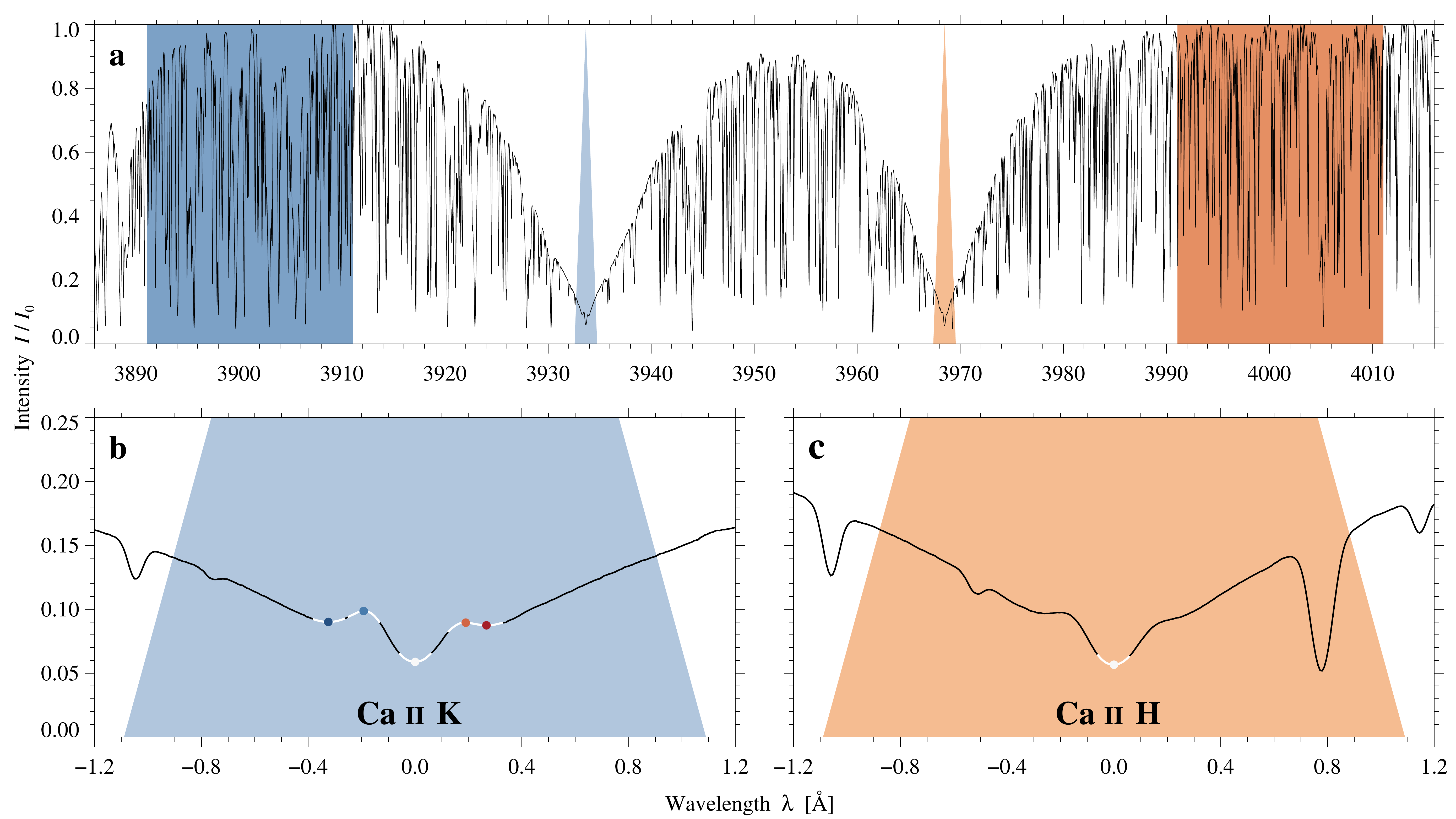}
\caption{(a) Average \cahk\ spectrum computed from 60 consecutive exposures 
    obtained by PEPSI/SDI on 2018 June~21. Details of the line fitting procedure are shown in the $\pm$1.2~\AA\ range around the cores of (b) the \cak\ and (c) the \cah\ chromospheric absorption lines. White segments indicate the range of parabola fits and colored bullets refer to local minima and maxima, i.e., K$_{1v}$ (\textit{dark blue}), K$_{2v}$ (\textit{light blue}), K$_3$ (\textit{white}), K$_{2r}$ (\textit{light red}), K$_{1r}$ (\textit{dark red}), and H$_3$ (\textit{white}). The 20~{\AA}ngstr\"om-wide wavelength bands of the blue and red pseudo continua are marked by light blue and light red rectangles, respectively. The blue and red triangular bands (FWHM = 1.09~\AA) are used in the computation of the $S$-index.}
\label{FIG01}    
\end{figure*}

A key science task of PEPSI/SDI is taking quasi-continuous observations of the Sun during one 11-year solar activity cycle. Whenever weather and technical conditions permit, PEPSI/SDI records up to 100 individual spectra per day, depending on the chosen wavelength range. This work is based on daily averages of spectra, which generally comprise 50\,--\,70 co-added individual spectra. An example of the strong chromospheric absorption lines \cahk\ is presented in \mbox{Figure~\ref{FIG01}{\ns}}. A more detailed description of this dataset and further processing steps are presented in Section~\ref{SEC3}. 

%-------------------------------------------------------------------------------
%   Figure 1 
%-------------------------------------------------------------------------------
%-------------------------------------------------------------------------------

%-------------------------------------------------------------------------------
%   Chromospheric Telescope
%-------------------------------------------------------------------------------

\subsection{Chromospheric Telescope}\label{SEC22}

The Chromospheric Telescope \citep[ChroTel,][]{Bethge2011, Kentischer2008}, operated by the Leibniz-Institut f\"ur Sonnenphysik (KIS) in Freiburg, Germany, is located at Observatory del Teide on Tenerife, Spain, where it performs synoptic observations of the Sun and delivers full-disk filtergrams in three strong chromospheric absorption lines, i.e., \cak\ 3933.7~\AA, H$\alpha$ 6562.8~\AA, and He\,\textsc{i}\,10830~\AA. ChroTel is by and large an automated instrument with an effective aperture of $D = 10$~cm. The sunlight is directed by a small turret and a system of mirrors to three narrow-band Lyot filters, thus obtaining consecutive observations in three wavelength channels with a cadence of three minutes. The diffraction-limited resolution of the telescope is according to the Rayleigh criterion $\alpha = 1.22 \lambda / D = 1.0^{\prime\prime}$ at \cak, $1.7^{\prime\prime}$ at H$\alpha$, and $2.7^{\prime\prime}$ at the near-infrared He\,\textsc{i} triplet. ChroTel employs a 2048$\times$2048-pixel CCD camera with a Kodak KAF-4320E detector. The quantum efficiency is still 35\% at \cak. Considering the image scale of about $1^{\prime\prime}$ pixel$^{-1}$, the \cak\ full-disk filtergrams are undersampled by a factor of two, i.e., the effective spatial resolution is either given by the detector or dominated by mediocre seeing conditions. 

%-------------------------------------------------------------------------------
%   Figure 2 
%-------------------------------------------------------------------------------
\begin{figure*}[t]
\centering
\includegraphics[width=0.876\textwidth]{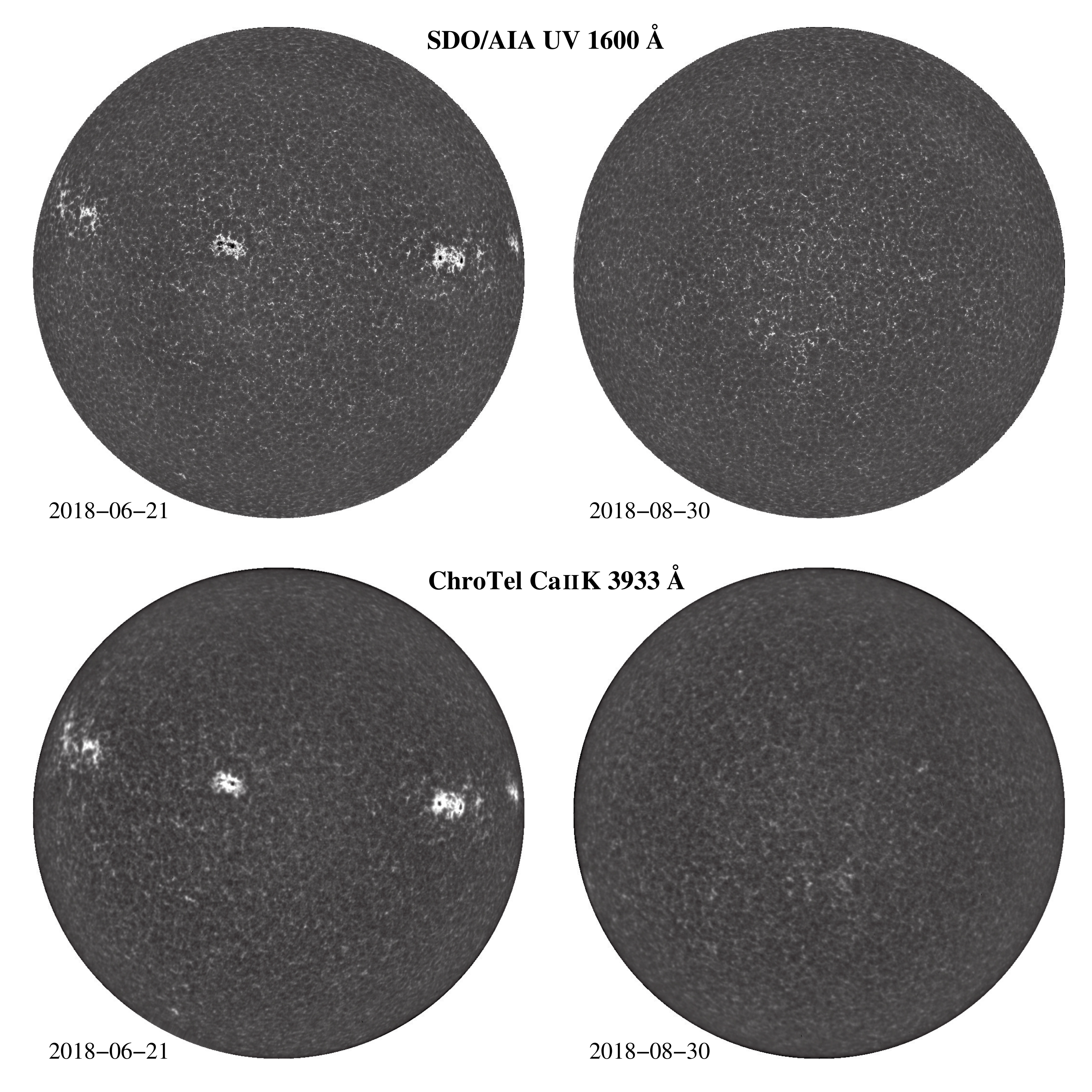}
\caption{Full-disk images from the AIA UV 1600~\AA\ channel (\textit{top}) 
    and the ChroTel \cak\ 3933~\AA\ Lyot filter (\textit{bottom}). Full-disk images representing moderate (2018 June~21) and low (2018 August~30) solar activity were selected according to the 2018 PEPSI/SDI $S$-index. A two-dimensional limb darkening function was subtracted from the full-disk images to enhance the contrast of the chromospheric fine structure. The brightest and darkest intensity values were clipped in the moderate-activity images, and the same intensity range was used for the low-activity images.}
\label{FIG02}    
\end{figure*}
%-------------------------------------------------------------------------------

The transmission curve of the \cak\ Lyot filter has a FWHM = 0.3~\AA, which is sufficiently broad to include the two emission reversal peaks at about $\pm$0.25~\AA\ around the line core (see Figure~\ref{FIG01}\ns). However, the Lyot filter can be configured for broader bandpasses (FWHM = 0.6~\AA\ and 1.2~\AA). The data reduction and correction procedures are automated and part of the operating software. Calibrated ChroTel filtergrams along with metadata are archived at the KIS Science Data Center\footnote{\href{http://sdc.leibniz-kis.de}{sdc.leibniz-kis.de}}, where they are publicly available in Flexible Image Transport System \citep[FITS,][]{Wells1981, Hanisch2001} format. The bottom row of Figure~\ref{FIG02}{\ns} displays two ChroTel \cak\ filtergrams observed at moderately high and low solar activity levels.

%-------------------------------------------------------------------------------
%   Atmospheric Imaging Assembly
%-------------------------------------------------------------------------------

\subsection{Atmospheric Imaging Assembly}\label{SEC23}

The Solar Dynamics Observatory \citep[SDO,][]{Pesnell2012} is a NASA mission and part of the \textit{Living with a Star} program, dedicated to monitoring solar magnetic activity throughout the activity cycle. Its instrument suite consists of three distinct instruments, i.e., the Helioseismic and Magnetic Imager \citep[HMI,][]{Schou2012}, the Atmospheric Imaging Assembly \citep[AIA,][]{Lemen2012}, and the Extreme Ultraviolet Variability Experiment \citep[EVE,][]{Woods2012}, where the first two instruments offer synoptic, high-cadence, multi-wavelength, full-disk observation of the Sun. During its decade-long operation since its launch in 2010 February~11, SDO collected vast amounts of data covering the ongoing Solar Cycle~24 from the preceding minimum in 2010 to present date, where another deep minimum just passed. 

This work is based on UV 1600~\AA\ narrow-band full-disk images obtained with one of AIA's four 20-centimeter Cassegrain telescopes. The telescope is equipped with a 4096$\times$4096-pixel CCD detector with a pixel size of 0.6\arcsec\ $\times$ 0.6\arcsec, delivering full-disk images with about 1.5\arcsec\ spatial and 12/24~s temporal resolution. The computation of the solar activity index utilizes long-integration UV images, a byproduct in the computation of a Background-subtracted Solar Activity Map \citep[BaSAM,][]{Denker2019}. Whereas a BaSAM  represents the temporal variation of the UV radiation in a time-series, the long-integration image establishes the background above which the variation is measured. 

The long-integration image significantly increases the photometric sensitivity of the UV images, while only slightly sacrificing spatial resolution because of the relatively slow temporal evolution of solar features. Each long-integration image represents the average of 300 individual UV images, after correction for differential rotation, covering a two-hour window centered on 12:00~UT. In addition, the long-integration images are normalized such that quiet-Sun disk center corresponds to unity, and a two-dimensional limb-darkening function is subtracted following the approach of \citet{Denker1999}. Remaining artifacts related to the field-dependent transmission profile of the UV filter were removed by modeling large-scale intensity variations across the solar disk by fitting low-order Zernike polynomials \citep{Shen2018}. Two normalized, limb-darkening corrected, long-integration UV 1600~\AA\ full-disk images are presented in the top row of Figure~\ref{FIG02}{\ns} to visualize moderate and low solar activity levels and to illustrate morphological similarities with the \cak\ full-disk images. The dataset in this study includes daily long-integration UV images between 2010 May~13 and 2019 September~30 with only a few interruptions caused by instrument anomalies and the three-week SDO eclipse seasons near the equinoxes. 

Figure~\ref{FIG02}{\ns} illustrates the difference between the quiet and the moderately active Sun. On 2018 June~21, the northern activity belt was covered with three $\beta$-class active regions (NOAA~12715, 12714, and 12713), which harbor sunspots, and a large plage region at the eastern limb. 

In contrast, no active or plage regions were present on the disk on 2018 August~30. However, the quiet Sun carries its own message, i.e., the enhanced network outlining the supergranular cells, which appear brightest at disk center in the UV 1600~\AA\ and \cak\ emissions. The strongest emission is visible near disk center and south of the equator.

%===============================================================================
%  SPECTRAL INDICES
%===============================================================================

\section{Spectral indices}\label{SEC3}

The broad-winged \cahk\ lines are a common feature in the absorption spectra of F, G, and K stars \citep{Linsky1970}. Spatially resolved observations of active region plage areas and network elements reveal significant emission enhancement in the \cahk\ line cores \citep[e.g.,][]{Skumanich1984, Zwaan1985, Ayres1986}. The predominant theory advocates that the enhancement is a result of magnetic heating \citep{Athay1970, Leenaarts2018}. \cahk-based indices and parameters are indicative of the variability of the chromosphere during the 11-year cycle \citep[e.g.,][]{Wilson1957, White1978, White1981}.

In the following, the PEPSI/SDI data reduction and derivation of the PEPSI/SDI $S$-index are presented. Furthermore, PEPSI data are probed for instrumental effects, and the sensitivity of the $S$-index is determined and validated by employing several solar references, i.e., spectral atlases. In addition to the $S$-index, nine \cak-line parameters \citep{Keil1998, Bertello2011} are computed to explore the full potential of the PEPSI/SDI high-resolution \cak\ spectra.

%-------------------------------------------------------------------------------
%   PEPSI/SDI S-index
%-------------------------------------------------------------------------------

\subsection{PEPSI/SDI \boldmath$S$-index}\label{SEC31}

In their seminal work, \citet{Vaughan1978} created a bridge to the original \cahk\ flux measurement reported in \citet{Wilson1968}. This became necessary because of the difference in operating the two HKP instruments, which were involved in the Mt.\ Wilson HK-monitoring project. In this work, we use as a reference the $S$-index as defined in \citet{Vaughan1978} for Mt.\ Wilson spectra. These four channels are indicated in Figure~\ref{FIG01}{\ns} following the exact wavelengths and passbands defined in \citet{Vaughan1978}.

The wide spectral range, simultaneously observed with PEPSI, easily includes the \cahk\ doublet and the adjacent continuum bands. The synoptic nature of PEPSI/SDI observations combined with the wide range and high spectral resolution motivated using Sun-as-a-star spectra in studying chromospheric activity variations \citep{Dineva2020}. 

All PEPSI spectra are processed by a dedicated data pipeline \citep[SDS4PEPSI,][]{Ilyin2000} including standard reduction procedures for high-resolution echelle spectra. A detailed account of all procedures and technical details is given in \citet{Strassmeier2018}. Bulk-processing of all raw echelle PEPSI data minimizes any inconsistencies, which can arise from non-uniform data treatment. The gravitational shift, radial velocity (RV), as well as any other velocity contribution from motion with respect to the observer are removed and the final wavelength scale is adjusted to the solar barycentric rest frame. Furthermore, PEPSI/SDI spectra are subjected to an absolute wavelength calibration, ensuring a wavelength accuracy on the order of meters per second. The resolution in the spectral window containing the \cahk\ doublet ${\cal R}$ is $\approx 220\,000$.

The PEPSI/SDI dataset contains 260 spectra in total (184 days in 2018 and 76 days in 2019). The PEPSI Sun-as-a-star spectra are well suited for examining the temporal evolution of the \cahk\ spectral indices over a period of about two years, whereby solar rotation plays a significant role in shaping the profile of the activity variations. An example of a PEPSI/SDI daily average spectrum is displayed in Figure~\ref{FIG01}{\ns}. The top panel shows the extended wavelength range of 124~\AA\ from 3889~\AA\ to 4013~\AA, which is required to compute the $S$-index.  

Each daily spectrum is a weighted average of up to 100 individual exposures. The purpose of the weighted average is to increase the SNR in the resulting spectrum. The weights are the inverse variance in each pixel, which is determined from the Poisson noise estimate derived from the original raw CCD image. Thus, the variance of the individual science-ready spectrum combines the error propagation in each step of the image processing. Finally, the weights are multiplied with the intensities at each pixel. Thus, the typical SNR is about 3400:1 for daily spectra, which is up to seven times that for a single exposure. Under the best observing conditions the SNR can even reach 5000:1. The solar flux obtained with the \mbox{PEPSI/SDI} telescope is normalized with respect to the solar spectrum atlas by \citet{Kurucz1984}, resulting in a flat, residual flux spectrum \citep{Strassmeier2018}. After interpolation, the initial wavelength increment of about 7~m\AA\ is sampled to a finer equidistant grid with increments of 1~m\AA, which facilitates a subsequent comparison with other reference spectra.

Changes in the PEPSI data reduction pipeline, which are related to the continuum correction, led to differences in the residual flux levels between the 2018 and 2019 normalized spectra. In particular, the difference is about 4.6\% and 2.1\% lower in the 2019 dataset blue and red pseudo-continua, respectively. The difference in the residual flux in the two line-cores is about 5.2\% and 4.4\% lower in the 2019 dataset, measured in a 1.09-{\AA}ngstr\"om range centered on the \cak\ and H line-core, respectively. This discrepancy was carefully taken into account for the time-series data presented in this study by carefully matching the 2019 to the 2018 data.

For the computation of the PEPSI solar $S$-index, we follow closely the recipe by \citet{Vaughan1978}. The index $S_\mathrm{PEPSI}$ is defined as the continuum normalized ratio between the integrated residual flux $F$ in the four aforementioned wavebands: 
\begin{equation}
S_\mathrm{PEPSI} = \alpha \times 8~ \frac{F_\mathrm{H} +F_\mathrm{K}}{F_\mathrm{R} + F_\mathrm{V}}. 
\label{EQN2}
\end{equation}
The constant value $\alpha = 2.4$ is recommended for the original Mt.~Wilson index $S_\mathrm{MW}$ by \citet{Vaughan1978} and later by \citet{Duncan1991}, as well as in the handling of contemporary normalized flux data \citep[e.g.,][]{Hall2007, Lovis2011, Schroder2012}. In the original HKP-2 measurements, the line-core passbands were exposed eight times longer than the continuum passbands. Therefore, we multiply the flux ratio by eight in Equation~(\ref{EQN2}). 

%-------------------------------------------------------------------------------
%   Figure 3
%-------------------------------------------------------------------------------
\begin{figure}[t]
\includegraphics[width=\columnwidth]{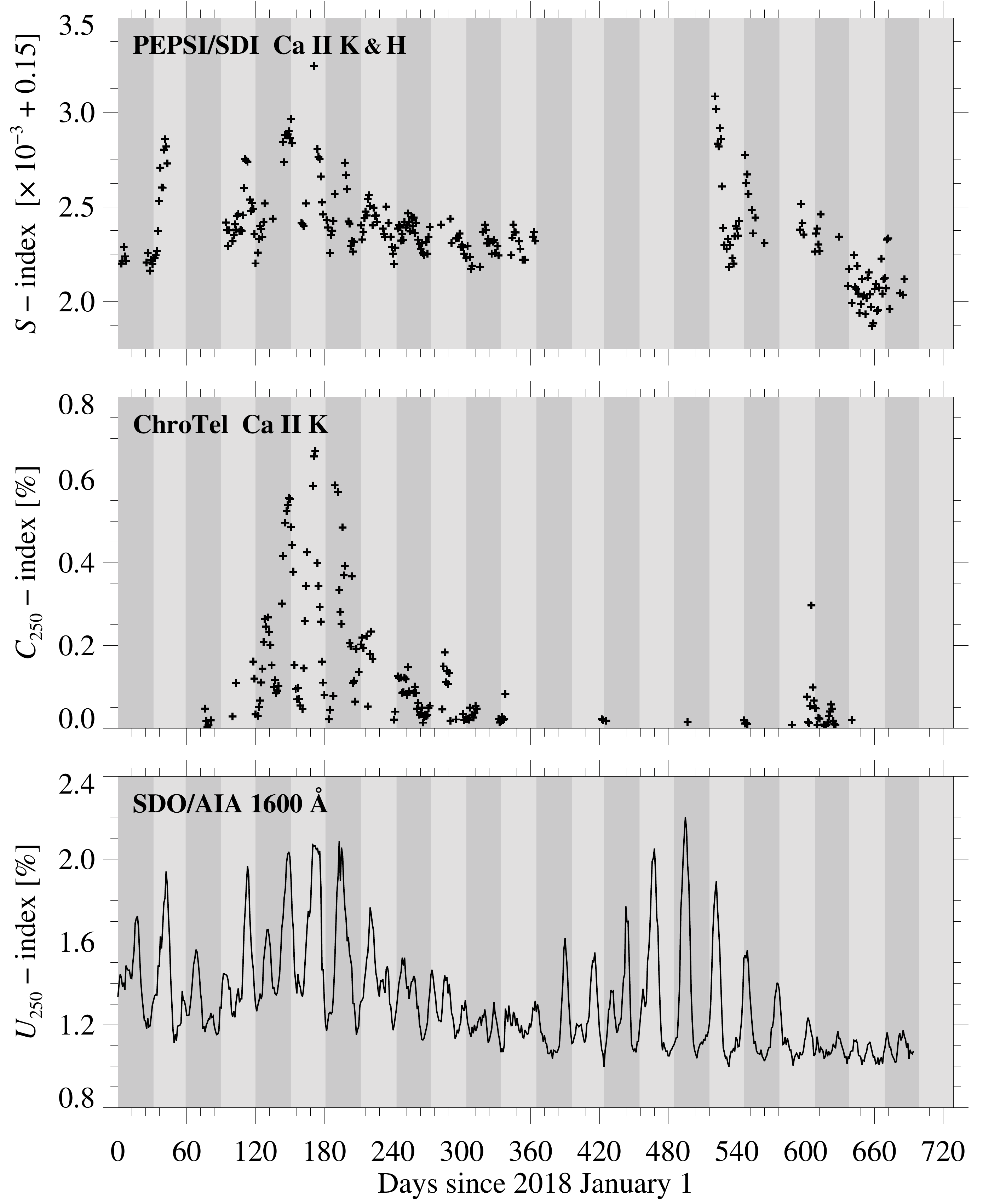}
\caption{Time-series of the PEPSI/SDI \cahk\ $S$-index (\textit{top}), ChroTel 
    \cak\ index $C_{250}$ (\textit{middle}), and AIA UV 1600~\AA\ index $U_{250}$ (\textit{bottom}). The alternating dark and light gray stripes denote one-month periods. The plus signs reflect the irregular data  coverage of the $S$- and $C_{250}$-indices, while the $U_{250}$-index has a continuous coverage.}
\label{FIG03}    
\end{figure}
%-------------------------------------------------------------------------------

The $S_\mathrm{PEPSI}$ time-series is displayed in the top panel of Figure~\ref{FIG03}{\ns}. The values vary between 0.152 and 0.153, with an average of $0.1522 \pm 0.0003$, where the standard deviation reflects the variation of solar activity rather than an error estimate. Note the offset and the scaling in the label of the ordinate. The difference between the 2018 and 2019 mean $S_\mathrm{PEPSI}$-values is of the magnitude of the standard deviation. This difference is so small that instrumental effects and the cross-calibration of the 2018 and 2019 data may leave an imprint. Nonetheless, the decline in $S_\mathrm{PEPSI}$ from 2018 to 2019 is in agreement with the deepening minimum of Solar Cycle~24. Furthermore, the index shows a clear indication of rotational modulation caused by bright chromospheric plages transiting the visible solar disk.  

The average $S_\mathrm{PEPSI}$-value during the declining phase of Solar Cycle~24 to the minimum is about 7\% lower than the $S$-value of 0.163 reported by \citet{Egeland2017} for the minimum period between Solar Cycles~23 and~24, which occurred in December 2008. In Table~1 of \citet{Egeland2017}, their results are placed in the context of previous studies \citep[i.e.,][]{Duncan1991, White1992, Baliunas1995, Radick1998, Hall2004}. The reported $S$-values are in the range 0.162\,--\,0.179 for the 2008 minimum, which can not just be attributed to measurement errors. \citet{Schroder2012} even obtained 8.6\% lower values than \citet{Egeland2017}, which are as low as $S$ = 0.150 on plage-free days. When approaching the December 2019 minimum, the $S_\mathrm{PEPSI}$-values come close to the extreme minimum values stated by \citet{Schroder2012}. In contrast, the average $S$-index in 2018 reported by \citet{Maldonado2019} based on HARPS spectra is 4.6\% higher than the respective average S$_\mathrm{PEPSI}$-index. Some discrepancies between the December 2008 and the December 2019 minima can be explained by poor sampling when taking the averages. Thus, solar cycle variations may significantly impact average $S$-values.  

Various steps, from obtaining the spectra to computing the indices, can contribute to discrepancies, which then accumulate to significant differences between the index values found in literature. In Section~\ref{SEC32}, we compare a PEPSI spectrum to various solar reference spectra. The fraction of the variance that can possibly arise from data handling is tested and discussed in Section~\ref{SEC33}. In Section~\ref{SEC33}, we demonstrate the intrinsic consistency and science capability of the $S_\mathrm{PEPSI}$ by discussing the rotation modulation of the two-year time-series.  

The purpose of the $S$-index, regardless of the instrument, is to create a scale of stellar magnetic activity, in which the Sun serves as a benchmark and stepping stone to other stars. Thus, any data processing procedure to compute the $S$-index has to ensure a close match with the $S_\mathrm{MW}$-index. This can be achieved, for example, via a set of standard stars \citep{Duncan1991}. Such a calibration is currently not available for PEPSI data. Therefore, we point out that the $S_\mathrm{PEPSI}$-index may still contain unresolved instrumental contributions.

%-------------------------------------------------------------------------------
%   Solar Spectra References
%-------------------------------------------------------------------------------

\subsection{Solar Spectral References}\label{SEC32}

Solar observation with high-spectral resolution obtained with Fourier Transform Spectrometer \citep[FTS,][]{Brault1985} at the McMath-Pierce Solar Telescope \citep{Pierce1964} at the Kitt Peak National Observatory in Arizona, U.S.A. have been extensively employed in the composition of high-fidelity spectral atlases. A variety of atlases are publicly available at the NSO Historical Archive.\footnote{\href{https://www.nso.edu/data/historical-archive/}{www.nso.edu/data/historical-archive}} The NSO FTS data contain observations of disk-integrated, disk-center, and sunspot spectra. The FTS construction and spectral restoration facilitates a spectral resolution ${\cal R}$ on the order of $10^6$. In addition, FTS observations provide high SNR data, virtually free of instrumental scattered light.   

The absolute disk-center intensity atlas \citep{Neckel1999} is widely used a quiet-Sun reference. It is compiled from McMath-Pierce FTS  observations by James W.\ Brault and colleagues, covering a small region near disk-center, and was obtained between 1980 and 1981 during a period of high solar activity. However, the disk-center intensities were chosen to reflect quiet-Sun conditions as close as possible. The spectral resolution is ${\cal R} \approx 350\,000$ near the \cahk\ doublet, with a wavelength increment of $\Delta\lambda = 2.0$~m\AA\ \citep{Doerr2016}.   

FTS observations of the same period, combining disk-integrated spectra and spectra of a region-of-interest (ROI), contributed to a solar flux atlas introduced by \citet{Kurucz1984}. \citet{Kurucz2006} presented a revised version of these solar flux data, which included a correction with a synthetic telluric spectrum and provided the associated residual flux. Both spectral references have high SNR ($\approx$~2500:1) and spectral resolution (${\cal R} \approx 400\,000$) in the blue spectral region with a wavelength increment of $\Delta\lambda = 4.6$~m\AA. More McMath-Pierce FTS data were obtained between 1981 and 1989, on days with considerably high magnetic activity, which resulted in an optical spectral atlas provided by \citet{Wallace2011}. The spectral resolution reaches up to ${\cal R} = 700\,000$ with a wavelength increment $\Delta\lambda = 2.4$~m\AA. 

First-light of PEPSI/SDI resulted in several Sun-as-a-star campaigns in 2015/16 \citep{Strassmeier2018}. The reduction procedures for daily spectra, as well as the main properties of the resulting spectra, are similar to the 2018/19 dataset employed in this work. Three observing days with the best data quality were selected for an initial analysis. Finally, three spectral profiles were computed as the weighted average of about 100 consecutive exposures. The integration time for a single spectrum was about 15~s, i.e., significantly longer than in the present dataset. \citet{Strassmeier2018} refers to these three datasets as PEPSI ``deep spectra''. 

Applying a scaling procedure described in \citet{Dineva2020} to the spectral atlases and the three PEPSI/SDI deep spectra facilitates a convenient and uniform cross-comparison with the daily-average PEPSI/SDI spectrum on 2019 October~20. This spectrum corresponds to the lowest level of solar activity with S$_\mathrm{PEPSI} = 0.152$. The cross-comparison is performed over the 130~\AA\ wavelength range shown in the top panel of Figure~\ref{FIG01}{\ns}. In addition, the central wavelengths for the \cahk\ lines are determined by a least-squares parabola fit. Owing to the different calibration procedure, a small blue- or a redshift may still be present in the different profiles and has to be corrected. Therefore, the photospheric singly ionized Fe\,\textsc{i} line at $\lambda$\,3932.63~\AA\ is used as a wavelength reference because telluric line blends are not available in the inner part of the \cak\ line. 

\begin{figure}[t]
\includegraphics[width=\columnwidth]{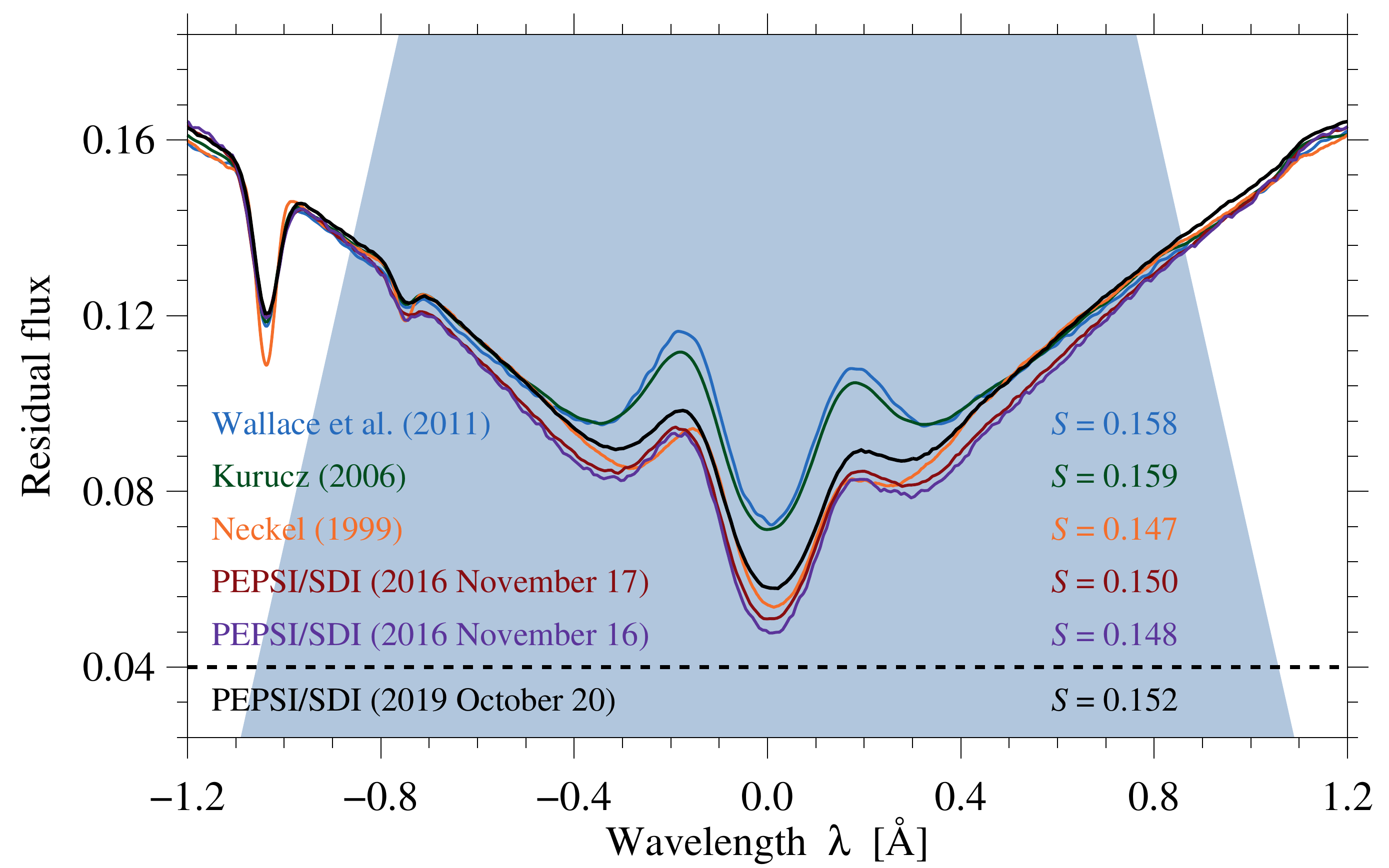}
\caption{Comparison of the \cak-line core features between the daily average
    PEPSI/SDI spectrum on 2019 October~20 (\textit{thick black}), the \citet{Wallace2011} atlas (\textit{light blue}), the \citet{Kurucz2006} atlas (\textit{green}), the \citet{Neckel1999} atlas (\textit{orange}), and two PEPSI ``deep spectra'' \citep{Strassmeier2018} on 2016 November~16 and~17. The black dashed line at a residual flux of 0.04 marks the basal quiet-Sun \cak-core depth. Color-coded $S$-indices for the atlas spectra are given on the right side of the plot panel. The low value in the case of \citet{Neckel1999} is due to the inherent differences between disk-integrated and disk-center atlases. }
\label{FIG04}    
\end{figure}

Figure~\ref{FIG04}{\ns} illustrates the result of the cross-comparison between three spectral atlases, two PEPSI deep spectra, and one daily-average PEPSI/SDI in the $\pm 1.2$~\AA\ wavelength range around the \cak\ line core. The light-blue area marks the triangular $F_\mathrm{K}$ bandpass, also indicated in Figure~\ref{FIG01}{\ns}b. The gray dashed line at the bottom of Figure~\ref{FIG04}{\ns} represents the low-limit, basal chromospheric \ca\ emission during an extreme minimum period identified by \citet{Livingston2007}. They speculate that such conditions may be representative for the Maunder Minimum \citep{Maunder1904}. The most pronounced differences in the profiles occur in the $\pm$0.5~\AA\ range around the central wavelength, affecting the emission reversal features and the line core. Certain variations can be attributed to disk-integrated versus ROI observations or high versus low solar activity, which are represented by emission enhancements \citep{Wallace2011, Kurucz2006} and basal emission \citep[PEPSI/SDI, PEPSI deep spectra,][]{Neckel1999}, respectively.  

%-------------------------------------------------------------------------------
%   Figure 4 
%-------------------------------------------------------------------------------
%-------------------------------------------------------------------------------

According to the analysis of the basal chromospheric flux by \citet{Schroder2012}, i.e., during minimum conditions or absence of surface activity, the corresponding $S$-index becomes as low as 0.150. The observations were obtained with data from the Heidelberg Extended Range Optical Spectrograph \citep[HEROS,][]{Hempelmann2005}, the main instrument of the Hamburg Robotic Telescope (HRT), between October~2008 and~July 2009, i.e., including extreme minimum conditions and the start of Solar Cycle~24. In comparison, the $S_\mathrm{PEPSI}$-index is 0.152 on 2019 October~20 (see black line profile in Figure~\ref{FIG04}{\ns}). The PEPSI deep spectra on 2016 November~16 and~17 result in $S$-index values of 0.148 and 0.150, respectively. These low values can be attributed to slightly different calibration procedures that were used at the time. While the two 20-\AA{ngstr\"om}-wide pseudo-continuum bands can be considered as a stable normalization factor, the $S$-index values depend highly on the line-core depths and emission reversals of the \cahk\ absorption lines. Thus, episodic events of low solar activity may be superposed on the solar activity cycle.

%-------------------------------------------------------------------------------
%   Validation of the S-index
%-------------------------------------------------------------------------------

\subsection{Validation of the \boldmath$S$-index}\label{SEC33}

A flat continuum is a common property of most contemporary disk-integrated, wide-range spectral datasets, i.e., the inherent slope of the solar radiation flux is lost. Due to heavy line blanketing and the wide wings of the \cahk\ doublet, the pseudo-continuum in the 122~\AA range is often below unity (see Figure~\ref{FIG01}{\ns}). Conversely, artifacts resulting from the composition of the cross-dispersed echelle spectra, the intensity calibration, and the normalization procedure may alter the residual flux. 
%-------------------------------------------------------------------------------
%   Figure 5 
%-------------------------------------------------------------------------------
\begin{figure}[t]
\includegraphics[width=\columnwidth]{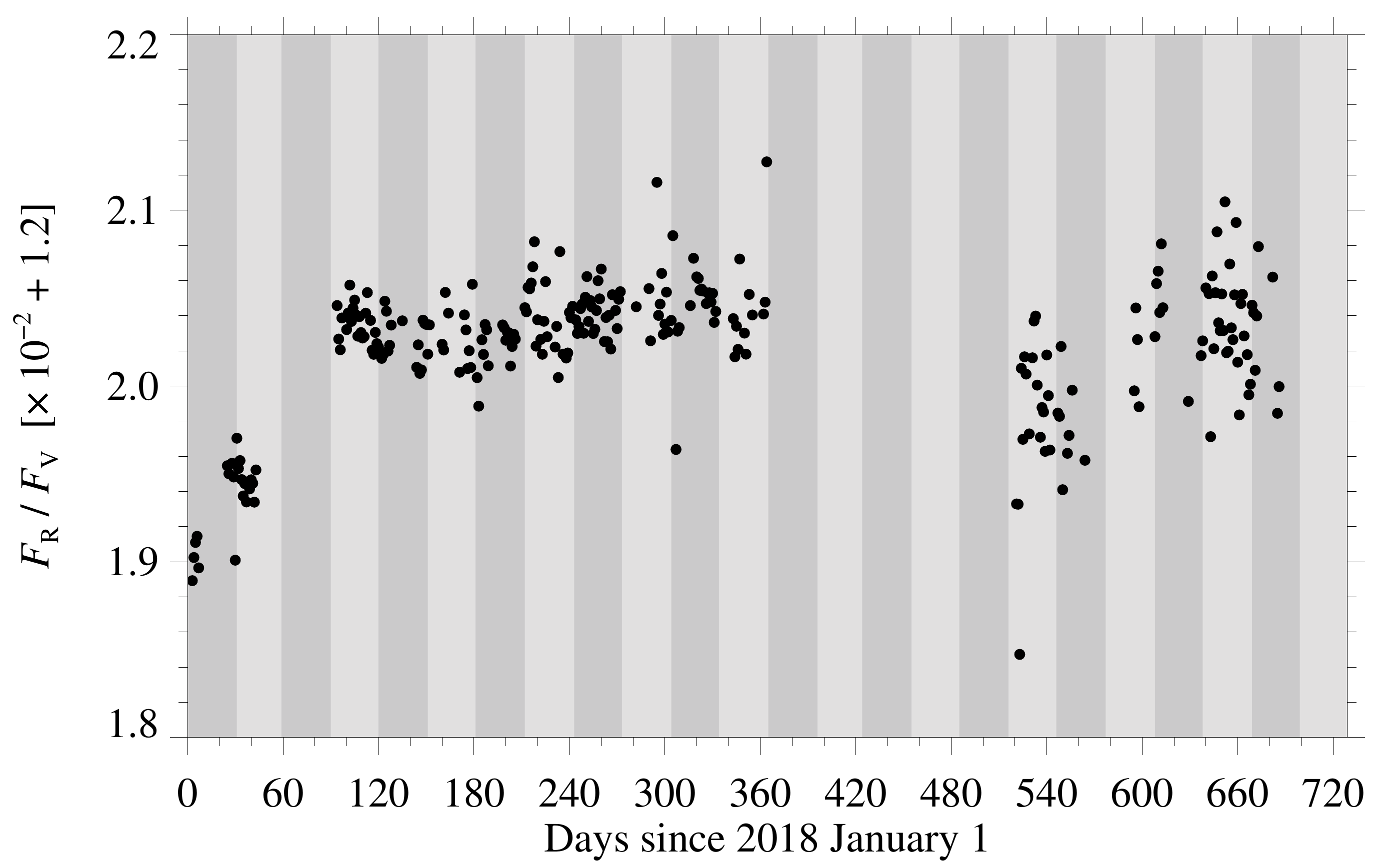}
\caption{Temporal evolution of the flux ratio of the red and blue PEPSI/SDI pseudo-continua $F_\mathrm{R}$ and $F_\mathrm{V}$, respectively. }
\label{FIG05}    
\end{figure}
%-------------------------------------------------------------------------------
The ratio between the integrated pseudo-continua $F_\mathrm{R}$ and $F_\mathrm{V}$ for the entire observing period is displayed in \mbox{Figure~\ref{FIG05}{\ns}} with an average value of 1.22. Comparing the top panel of \mbox{Figure~\ref{FIG03}{\ns}} to \mbox{Figure~\ref{FIG05}{\ns}} shows that the ratio varies with time but without a clear dependence on solar activity and that the  rotational modulation is absent. Thus, instrumental and environmental effects may introduce an error of about 0.2\%. Such effects are even lower for the time period 2018 April\,--\,December, but also outliers exist, for example in 2019 June. These variations must be taken into account when discussing the robustness and accuracy of the $S_\mathrm{PEPSI}$-index time-series. 

The off-axis design of the SDI telescope in combination with an integration sphere ensures that observations are free of guiding jitter. Furthermore, the elaborate wavelength calibration procedures and consecutive daily averaging of single exposures result in a spectrum closely resembling an atmosphere at rest. That is, any global or local oscillation signals are smeared by the averaging. The remaining shifts of the \cah\ and \cak\ line-core positions $\lambda_{\mathrm{K}\scriptscriptstyle3}$ and $\lambda_{\mathrm{H}\scriptscriptstyle3}$ with respect to the central wavelengths are related to chromospheric variations during the 11-year activity cycle. The temporal evolution of $\lambda_{\mathrm{K}\scriptscriptstyle3}$ is discussed in Section~\ref{SEC34}, among other \cak-line parameters. 

Selected PEPSI/SDI daily-average spectra, representing high, average and low $S_\mathrm{PEPSI}$-indices, are subjected to several numerical experiments, which were designed to test the self-consistency of the 2018 and 2019 datasets. These experiments aim to quantify any measurement errors, which may lead to discrepancies between the $S_\mathrm{PEPSI}$-index and other $S$-index values presented in the literature. Two major cases are considered, i.e., inaccurate placement in wavelength of the four integration passbands and modifications of the pseudo-continuum slope. 

In the first experiment, the two rectangular passbands $F_\mathrm{R}$ and $F_\mathrm{V}$ are shifted up to $\pm 0.5$~\AA\ with respect to each-other.  Deviations from the initial $S_\mathrm{PEPSI}$-value are less than 1\% and independent of the direction of the displacements. The same procedure is applied to the two triangular bandpass, $F_\mathrm{K}$ and $F_\mathrm{H}$, centered on the two line cores. In this case, the largest increase in the $S_\mathrm{PEPSI}$-value of about 18\% is associated with the largest displacement of the triangular passband. To ensure that the $S_\mathrm{PEPSI}$-value deviates less than 1\% from the initial value, the wavelength displacement must be smaller than 0.12~\AA. Thus, the $S$-index is very sensitive to zero-point offsets of the triangular passbands. Moreover, a change in the FWHM of the triangular passband has a significant effect on the $S_\mathrm{PEPSI}$-index. Already a $\pm10$~m\AA\ difference in the FWHM results in a $\pm 1.2$\% change in the $S_\mathrm{PEPSI}$-values. 

The seconds experiment concerns the trend removal applied to the pseudo-continuum. The trend in a spectrograph profile may be related to various technical issues that can arise during the observations or to the choice of data reduction procedures. In the experiment, the selected daily-average PEPSI/SDI spectra are multiplied by a linear trend, where the mean value is unity and the slope varies such that the endpoints cover the range 0.9\,--\,1.1. The Planck function, for an effective temperature $T_\mathrm{eff}$~=~5777~K serves as a benchmark for the theoretical slope of the solar flux in the given wavelength interval. The largest change occurs in the $F_\mathrm{R}$ and $F_\mathrm{V}$ values, which vary by up to $\pm 8.2$\%. However, the $S_\mathrm{PEPSI}$-values are much more stable and only change by less than 1\%. 

Changing solar activity, which strongly affects the \cahk\ line cores, was not considered in the numerical experiments. However, the line cores are the locations, where the $S_\mathrm{PEPSI}$-index values are most sensitive to measurement errors. Therefore, proper data calibration is important to avoid biases in $S$-index time-series. 

In Section~\ref{SEC31}, the discrepancies between the $S_\mathrm{PEPSI}$-index time-series and others presented in literature \citep{Egeland2017, Schroder2012, Maldonado2019} are up to about 7\%. Considering the results from the numerical test, the $S_\mathrm{PEPSI}$-indices are consistent with declining chromospheric activity, approaching the minimum between Solar Cycles~24 and~25. A further improvement in calibrating the \mbox{PEPSI/SDI} spectra can be expected from comparison with a sample of Mt.\ Wilson standard stars, which will also facilitate a more direct comparison with other datasets. 

The $S_\mathrm{PEPSI}$-index time-series plotted in the top panel of Figure~\ref{FIG03}{\ns} clearly shows rotational modulation. For example, the large peak in 2018 June~21 (three $\beta$-class active regions) and the two more moderate peaks in 2018 February~11 (one $\beta$-class active region) and 2018 April~22 (one $\alpha$-class and one $\beta$-class active region) indicate the disk passage of large active regions with surrounding bright plages. These are followed by many smaller peaks with gradually reduced $S_\mathrm{PEPSI}$-values, which trace the declining phase of Solar Cycle~24. Unfortunately, a large gap in the PEPSI/SDI observations of almost five months between 2019~January and May prevented recording data during a period of enhanced solar activity (see other activity indices in the middle and bottom panels of Figure~\ref{FIG03}{\ns}). The $S_\mathrm{PEPSI}$-values for 2019 show two consecutive peaks in June towards the end of this time period. The remainder of 2019 is characterized by a basal level of the $S_\mathrm{PEPSI}$-index, which is even lower compared to 2018 and does not show any clear indication of rotational modulation. Possible explanations included the aforementioned changes in the PEPSI/SDI data reduction pipeline (see Section~\ref{SEC31}), poorer sampling in 2019, and seeing/environmental conditions, which may obfuscate rotational modulation. Synoptic observations from space (see bottom panel of Figure~\ref{FIG03}{\ns}) still show rotational modulation even at extremely low activity levels, where   
only the enhanced chromospheric network plays a role in sustaining the basal chromospheric emission. 

%-------------------------------------------------------------------------------
%   Figure 6 
%-------------------------------------------------------------------------------
\begin{figure}[t]
\includegraphics[width=\columnwidth]{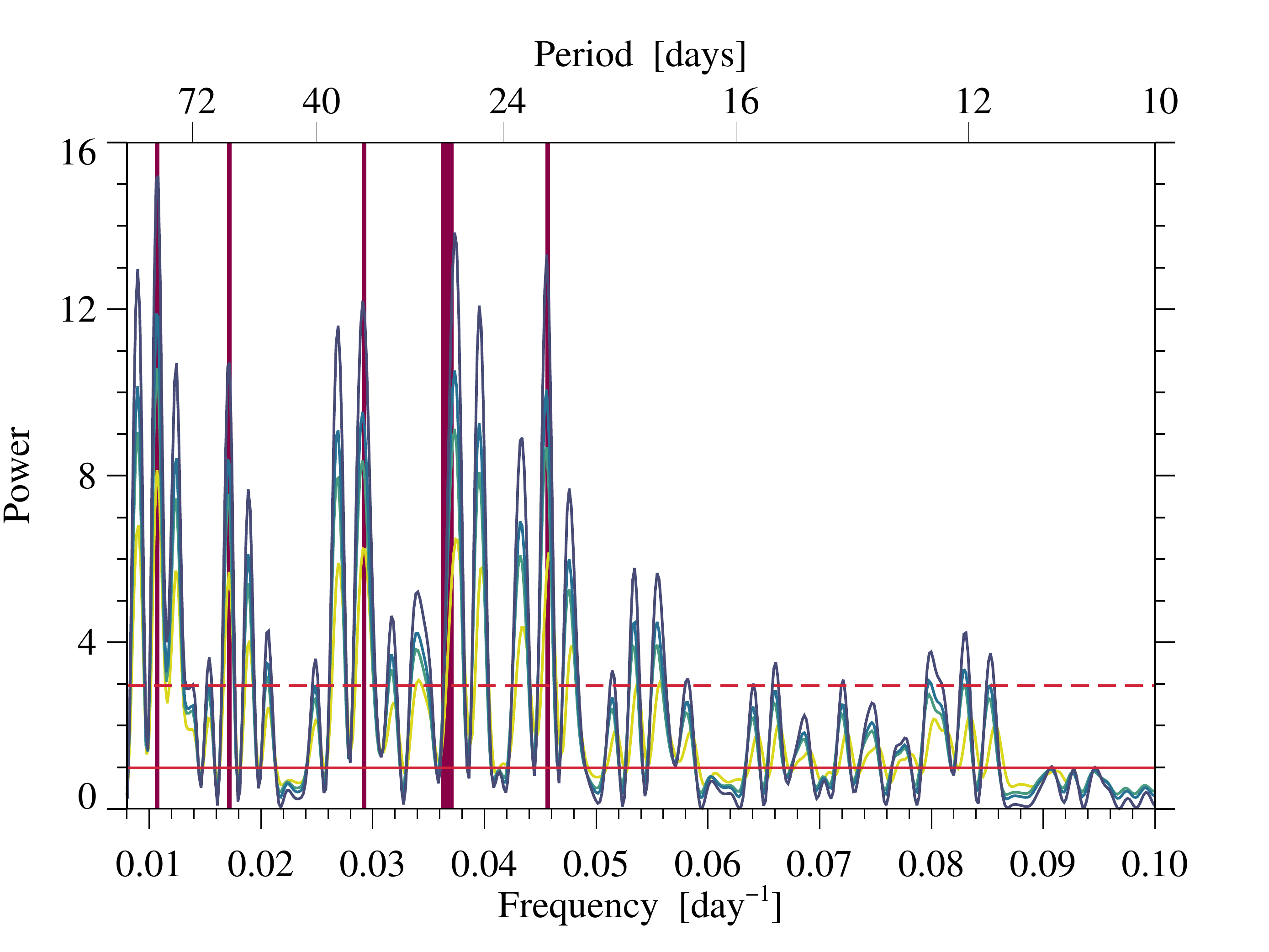}
\caption{Lomb-Scargle periodogram (\textit{dark blue} curve) of the PEPSI/SDI 
    $S$-index time-series for 2018 and 2019. Three periodograms (\textit{light blue}, \textit{cyan}, and \textit{yellow} curves) refer to validation tests, when only three-quarters, two-thirds, and one-half of the data points were used. A noise estimate (horizontal \textit{red} line) is based on the average of 1000 samples, where the $S$-index values were randomly shuffled. Power peaks that are three times this value (horizontal dashed \textit{red} line) can be considered significant. The Carrington rotation period of 27.3~days (thick vertical \textit{purple} line) is given for reference. Other significant peaks (vertical \textit{purple} lines) refer to periods of 21.9, 34.2, 58.3, and 93.5}~days.
\label{FIG06}    
\end{figure}
%-------------------------------------------------------------------------------

In another validation step, the qualitative description of the rotational modulation of the $S_\mathrm{PEPSI}$-index is complemented by a quantitative analysis. In the case of unevenly sampled data, such as the $S_\mathrm{PEPSI}$-index, the preferred method of estimating the power spectrum is the Lomb-Scargle periodogram \citep{Lomb1976,Scargle1982}. 

The Lomb-Scargle periodogram, which was computed from the $S_\mathrm{PEPSI}$-index time-series, is presented in Figure~\ref{FIG06}{\ns}. The dark blue curve indicates the power spectrum estimated from the entire dataset. Three additional sets of power estimates were computed from three-quarters, two-thirds, and one-half of the data points to validate the significance of the periodic signatures. Each curve represents an ensemble average based on 1000 samples, where the $S_\mathrm{PEPSI}$-index values were randomly selected. The horizontal red line provides a noise estimate, which is based on 1000 samples of randomly shuffled $S_\mathrm{PEPSI}$-index values. A robust criterion for the significance of a peak in the power spectrum is when the peak exceeds three times the noise estimate, which is further indicated by the dash-dotted red line. The large gap between the 2018 and 2019 data results in high power at low frequencies. Therefore, the abscissa only starts at a frequency of 0.01~day$^{-1}$.

The Carrington rotation period of 27.3~days is marked with a thick purple vertical line. The immediately adjacent power peak corresponds to a period of 26.7~days and is accompanied by a smaller power peak at higher frequencies. In addition, other peaks with significant power are present with periods of 21.9, 34.2, 58.3, and 93.5 days, which can be related to (sub-)harmonics considering accuracy and precision of the measurements. The presence of side band may be attributed to the unevenly sampled data and to the unbalanced presence of activity features in the northern and southern hemisphere.

In summary, robustness and accuracy of the $S_\mathrm{PEPSI}$-index were validated by a set of numerical experiments, which determined that the only significant deviations in the $S_\mathrm{PEPSI}$-index values may occur due to inconsistencies in the $F_\mathrm{R}$ and $F_\mathrm{V}$ passbands. 

Furthermore, the power spectrum analysis allowed us a self-consistency check, uncovering the Carrington period from the unevenly sampled $S_\mathrm{PEPSI}$-index time-series -- even during declining solar activity towards the end of Solar Cycle~24.

%-------------------------------------------------------------------------------
%   Figure 7
%-------------------------------------------------------------------------------
\begin{figure*}[t]  
\includegraphics[width=\textwidth]{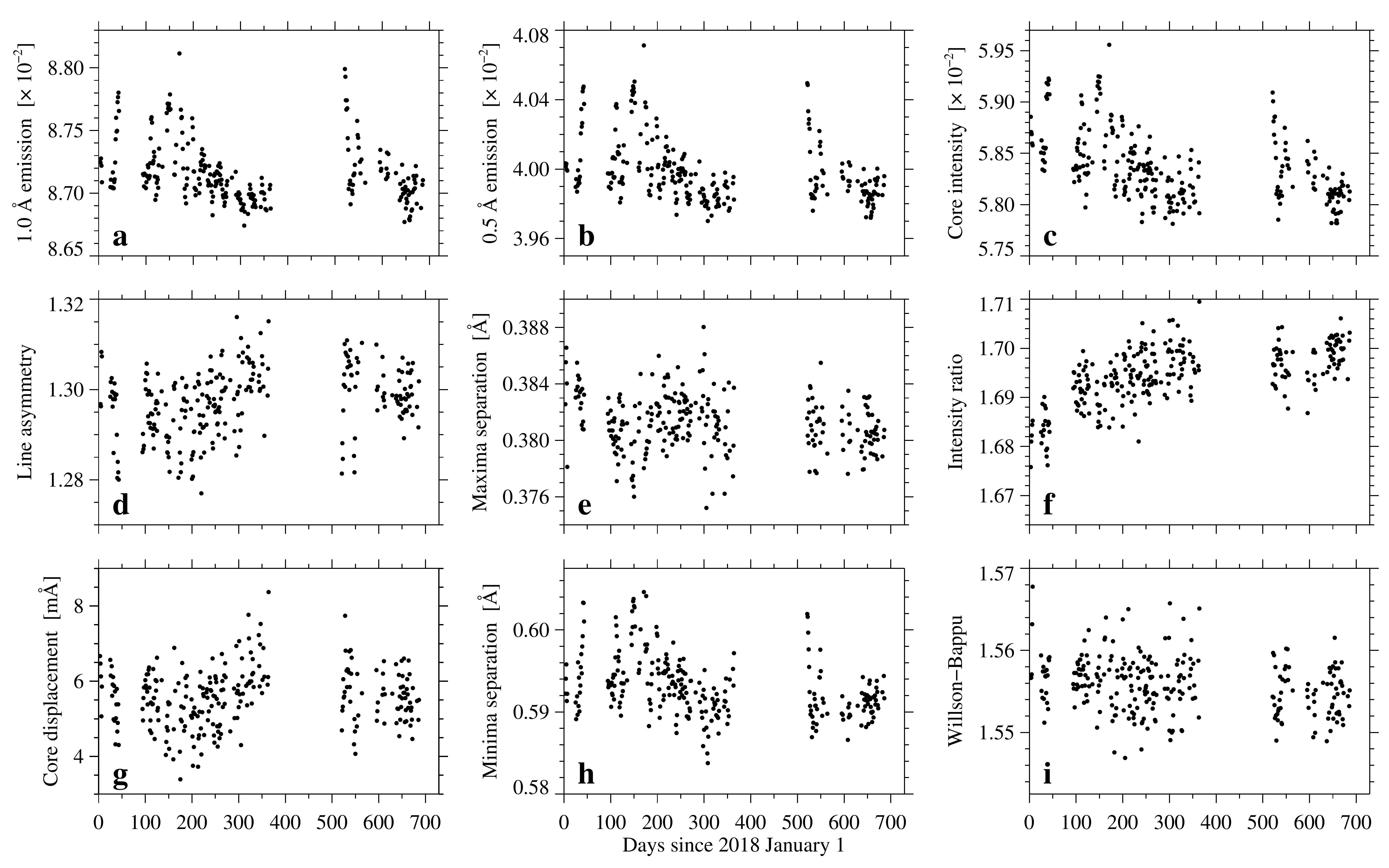}
\caption{Time-series of the \cak-line parameters: (a) integrated emission in a    
    1.0-{\AA}ngstr\"om-wide passband, (b) integrated emission in a    
    0.5-{\AA}ngstr\"om-wide passband, (c) K$_{3}$ intensity, (d) line asymmetry, (e) separation of K$_{2v}$ and K$_{2r}$ maxima, (f) K$_{2v}$ and K$_{3}$ intensity ratio, (g) line-core displacement with respect to the \cak\ central wavelength, (h) separation of K$_{1v}$ and K$_{1r}$ minima, and (i) Wilson-Bappu parameter. }
\label{FIG07}    
\end{figure*}
%-------------------------------------------------------------------------------

%-------------------------------------------------------------------------------
%   Ca II K-line parameters
%-------------------------------------------------------------------------------

\subsection{\cak-line parameters}\label{SEC34}

The \cahk\ lines have similar profiles, although the \cak\ line has a more clearly defined emission reversal, and notably so during low activity periods (Figure~\ref{FIG01}\ns). This characteristic inspired \citet{White1978} to introduce a set of parameters within a 1-{\AA}ngstr\"om-wide passband around the \cak\ line core. The five profile features relative to the \cak-line parameters are marked with color-coded bullets in Figure~\ref{FIG01}{\ns}b. The most extensive record of these features, which consists of observations obtained at NSO's Sacramento Peak Observatory (NSO/SP), covers several solar activity cycles from 1974~November to 2015~October.\footnote{\href{https://lasp.colorado.edu/lisird/data/cak/}{lasp.colorado.edu/lisird/data/cak/}} This program was continued by NSO's SOLIS facility at Kitt Peak National Observatory (KPNO).\footnote{\href{https://solis.nso.edu/0/iss/iss_timeseries.html}{solis.nso.edu/0/iss/iss\_timeseries.html}}

The PEPSI/SDI data bridge the time when SOLIS was relocated from KPNO in 2014, temporarily moved to Tucson until 2018, and is now permanently installed at BBSO. Furthermore, the \cak-line parameters complement the $S_\mathrm{PEPSI}$-index, linking spectral features to variations of solar activity. 

The wavelength position and residual flux values are determined by least-squares, second degree polynomial fitting in the range indicated by white color in Figure~\ref{FIG01}{\ns}b. For consistency with the established nomenclature, the term ``intensity'' is used instead of residual flux, when referring to \cak-line parameters.

The 1-{\AA}ngstr\"om-integrated emission parameter is essentially the equivalent width measured in a 1-{\AA}ngstr\"om-wide passband centered at $\lambda_{\mathrm{K}\scriptscriptstyle3}$ (Figure~\ref{FIG07}{\ns}a). Hence, this parameter reflects the variation of the K$_3$ intensity with solar activity. According to line-formation theory, the 1-{\AA}ngstr\"om-wide passband is contaminated by photospheric emission from the line wings. Another parameters, the 0.5-{\AA}ngstr\"om-integrated emission (Figure~\ref{FIG07}{\ns}b), i.e., the equivalent width, isolates strictly the chromospheric portion of the \cak\ emission \citep{Linsky1978, Pevtsov2013}. 

Other intensity-based parameters are the core intensity at $\lambda_{\mathrm{K}\scriptscriptstyle3}$  (Figure~\ref{FIG07}{\ns}c), the line asymmetry (Figure~\ref{FIG07}{\ns}d), and the intensity ratio (Figure~\ref{FIG07}{\ns}f). The intensity ratio, which is taken between the blue emission reversal maximum K$_{2v}$ and the absorption minimum K$_{3}$, measures the emission reversal strength. The line asymmetry is defined as the intensity computed by dividing the intensity difference between K$_{2v}$ and K$_{3}$ and the intensity difference between K$_{2r}$ and K$_{3}$. 

The wavelength-based \cak-line parameters include the $\lambda_{\mathrm{K}\scriptscriptstyle3}$ displacement with respect to the \cak\ central wavelength at 3932.67~\AA\ (Figure~\ref{FIG07}{\ns}g), the separation between the two maxima K$_{2v}$ and K$_{2r}$ (Figure~\ref{FIG07}{\ns}e), the separation between the two minima K$_{1v}$ and K$_{1r}$ (Figure~\ref{FIG07}{\ns}h), and the Wilson-Bappu parameter
\begin{equation}
W = \log \left[76.28\, (\lambda_{\mathrm{R}} - \lambda_{\mathrm{V}}) \right],
\label{EQN3}
\end{equation}
where $\lambda_{\mathrm{R}}$ is the wavelength position of the point half-way between K$_{1r}$ and K$_{2r}$, and $\lambda_{\mathrm{V}}$ is measured between K$_{1v}$ and K$_{2v}$ (Figure~\ref{FIG07}{\ns}i). The Wilson-Bappu parameter was introduced by \citet{Wilson1957} as a measure of the enhanced \cak\ line-core emission expressed as a logarithmic width. 

Time-series of the hereto described nine \cak-line parameters are displayed according to their variation trend, i.e., their decrease or increase with respect to solar activity. Since long-term records of these parameters are available from other instruments, their correlation with respect to other activity indices, e.g., with the relative sunspot number (SSN),  is known \citep{Livingston2007,Bertello2016}. 

All parameters displayed in the top row of \mbox{Figure~\ref{FIG07}{\ns}} show a similar temporal variation as the $S_\mathrm{PEPSI}$-index plotted in the top panel of \mbox{Figure~\ref{FIG03}{\ns}}. A distinct pattern of rotational modulation is evident in the two integrated emission parameters, and to a lower degree in the K$_3$ intensity. Indeed, Lomb-Scargle periodograms (not shown) for time-series of the 0.5- and 1.0-{\AA}ngstr\"om-integrated emission display a distinct power peak matching the Carrington rotation period. In addition, the amplitude of the variation increases between 2.0\% to 3.0\% in Figure~\ref{FIG07}{\ns}a--c even at times of low activity.

The three parameters displayed in the middle row of Figure~\ref{FIG07}{\ns} exhibit opposite trends compared to the ones in the top row. In quiet-Sun areas, the K$_{2v}$ peak in \cak\ profiles is usually stronger than the K$_{2r}$ peak \citep{Carlsson1997}. Hence, the line appears asymmetric \citep{White1981}.  This effect becomes negligible with rising solar activity. Spatially resolved plage spectra show almost symmetrical emission features. A weak K$_{2r}$ peak indicates a low number or the absence of bright plages on the solar surface. Therefore, the line-asymmetry increases around the activity cycle minimum.

The wavelength separation between the two maxima shows little deviation from its mean value, although it increases when surface activity is low. According to \citet{Bjorgen2018} the larger separation, i.e., a broader emission profile, is caused by heating above the temperature minimum. 

The intensity ratio displayed in Figure~\ref{FIG07}{\ns}f also increases when approaching minimum conditions at the end of Solar Cycle~24. Although the intensity in both K$_{2v}$ and K$_3$ decreases, this happens at a different rate.

%-------------------------------------------------------------------------------
\begin{figure*}[t]
\includegraphics[width=\textwidth]{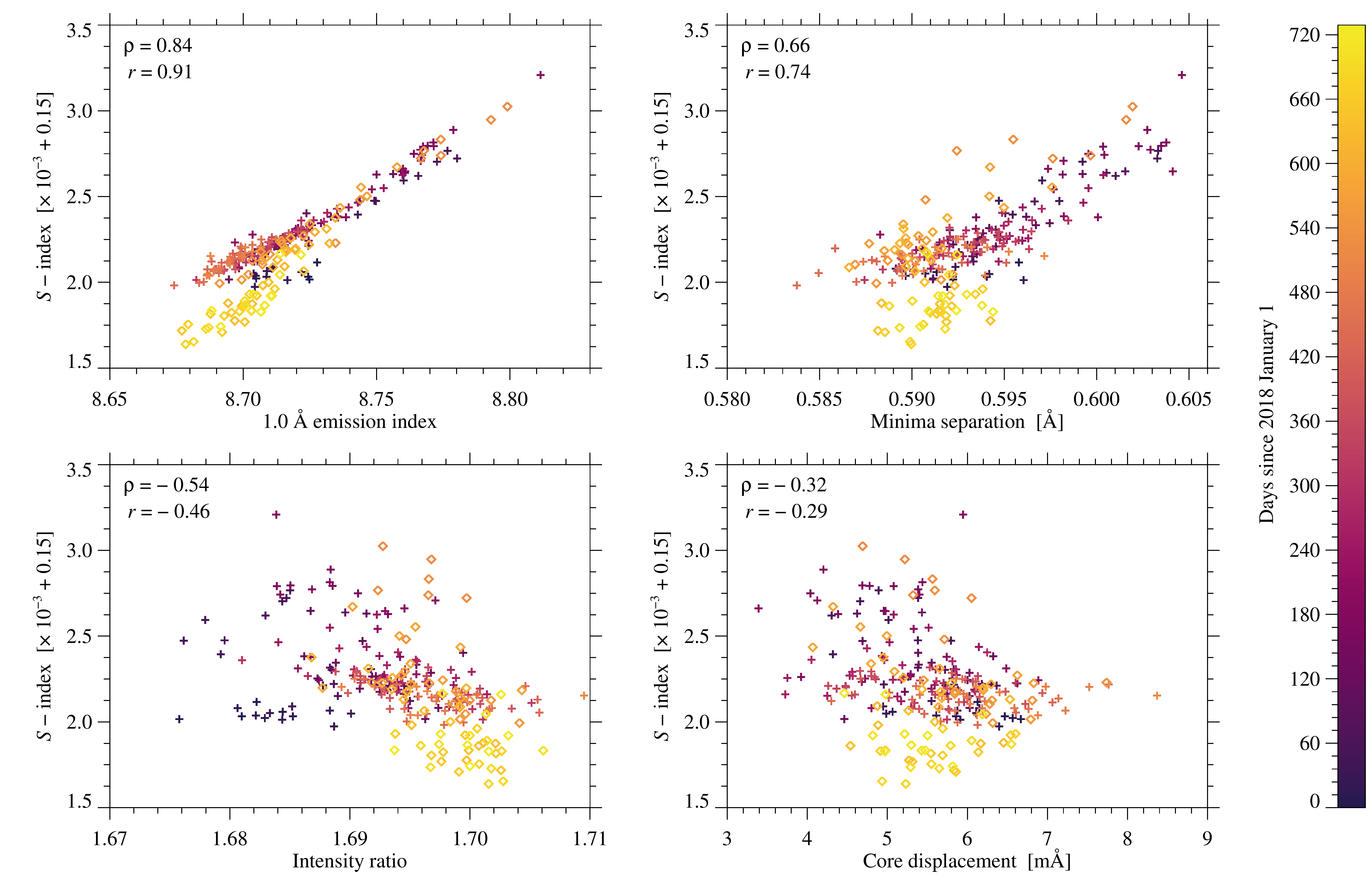}
\caption{Scatter plots of the $S_\mathrm{PEPSI}$-index vs.\ the integrated
    emission in a 1.0-{\AA}ngstr\"om-wide passband, the K$_{2v}$ and K$_{3}$ intensity ratio, and the line-core displacement with respect to the \cak\ central wavelength. Spearman's rank-order and Pearson's linear correlation coefficients $\rho$ and $r$ are given in the top-left corners of each panel, respectively. The ``plasma'' color table indicates the chronological order of the data points starting on 2018 January~1. In addition, two types of symbols indicate the data point which belong to 2018 (\textit{crosses}) and 2019 (\textit{diamonds}) datasets.}
\label{FIG08}    
\end{figure*}
%-------------------------------------------------------------------------------

The last three \cak-line parameters are displayed in the bottom row of Figure~\ref{FIG07}{\ns} and show a mixture of increasing and decreasing trends. The K$_3$ wavelength is redshifted with respect to the central wavelength. The less magnetic activity is observed on the solar disk, the larger the redshift. Through comparison between synthetic and spatially averaged spectra, \citet{Bjorgen2018} demonstrated that the $\lambda_{\mathrm{K}\scriptscriptstyle3}$ is a powerful diagnostics of vertical velocities in the chromosphere. 

Figure~\ref{FIG07}{\ns}h presents the K$_1$ minima separation, which mirrors the temporal evolution of the K$_2$ maxima separation. In spatially resolved spectra of active region plages, the minima separation increases. Thus, a decrease of the minima separation indicates fewer active region plages on the visible disk. Furthermore, the rotation modulation pattern is well defined in the minima separation. Indeed, power spectrum estimates computed from these time-series reveal a well-defined power peak near the Carrington rotation period. Finally, the Wilson-Bappu parameter is stable around its mean value.                                                                                                                                                                                        
The hereto discussed results are consistent with the measurements presented by \citet{White1981} and \citet{Livingston2007} for the low activity phase of the few observed solar cycle minima between 1974 and 2006. Although the results agree in magnitude, each minimum has its own basal level of activity, which is not necessarily the same as in the previous or the following minima. 

While the various trends of the PEPSI/SDI \cak-line parameters agree with the findings of \cite{Bertello2011}, scaling factors between the SOLIS/ISS and PEPSI/SDI data have to be considered. The SOLIS/ISS observations started in 2006~December and comprise a much longer period than the PEPSI/SDI data. \citet{Dineva2020a} demonstrated that the core of the \cak-line obtained with PEPSI/SDI is deeper compared to SOLIS/ISS for the same day. This difference in depth affects the intensity-based parameters. In addition, \cite{Bertello2011} applied a more elaborate fitting method to the \cak-line core, using a decomposition into Fourier components before integrating the emission in the fitted range. 

Four parameters with strong (anti-)correlations, indicated by the Spearman's and Pearson's correlation coefficients $\rho$ and  $r$, respectively, are displayed in Figure~\ref{FIG08}{\ns}, where color and shape of the markers indicate the time of observations. The top-left panel of Figure~\ref{FIG08}{\ns} shows a scatter plot of the $S_\mathrm{PEPSI}$-index vs.\ the 1.0-{\AA}ngstr\"om-integrated emission. The latter is closely related to the $F_\mathrm{K}$ value (see Equation~\ref{EQN2}). Therefore, these two time-series exhibit the strongest correlation. In the neighboring panel, the correlation between the $S_\mathrm{PEPSI}$-index and the minima separation is significantly smaller, although it is the second strongest. In contrast to the 1.0-{\AA}ngstr\"om-integrated emission, the minima separation seems to approach a lower limit of about 0.585~m\AA, which is also evident in Figure~\ref{FIG07}{\ns}h and represents likely the basal level of chromospheric activity. 
 
The parameters, which show negative or inverse correlations with the $S_\mathrm{PEPSI}$-index, are displayed in the two bottom panels of Figure~\ref{FIG08}{\ns}. The relation between the intensity ratio and chromospheric optical depth \citep{Skumanich1967, White1981} implies that when the optical depth increases in the chromosphere, the Sun-as-a-star \cak\ emission decreases, hence the negative correlation coefficients. However, this relationship is not very strong. Finally, the inverse correlation between the $\lambda_{\mathrm{K}\scriptscriptstyle3}$ displacement and the $S_\mathrm{PEPSI}$-index is even weaker as shown in the bottom-right panel of Figure~\ref{FIG08}{\ns}.

All scatter plots in Figure~\ref{FIG08}{\ns} are characterized by a clustering of the low activity values (yellow diamonds), which represent late 2019, where the lowest $S_\mathrm{PEPSI}$-index values were measured. One obvious reason is the absence of any plage regions, resulting in an extremely weak rotational modulation (see Figures~\ref{FIG03}{\ns}  and \ref{FIG07}{\ns}). Since the $S$-index is a product of the emission in the \cahk\ line cores, only the network continues to contribute. This may also explain the two-branch structure at low activity, most prominently seen in the top-left panel of Figure~\ref{FIG08}{\ns}.

%===============================================================================
%   AREA AND EXCESS INTENSITY INDICES
%===============================================================================

\section{Area and excess intensity indices}\label{SEC4}

Bright features related to high concentrations of magnetic fields, i.e., active region plages and chromospheric network, govern the variations in the \cahk\ emission in the chromosphere. During high-activity phases of the solar cycle, excessively bright active region plages outshine the enhanced network. However, during solar minimum conditions, the enhanced network is the sole contributor to the basal \cahk\ emission. 

Using disk-resolved \cak\ filtergrams, activity indices are computed, which monitor the excess emission and areal coverage of plage and network, thus, quantifying chromospheric surface morphology and activity. In addition, a similar index pair is derived from full-disk images recorded at UV wavelengths. The main difference between \cak\ and UV indices is that the latter refer to lower chromospheric heights and are less susceptible to canopy effects \citep{Krijger2001}. 

In the following, processing of disk-resolved images and computing excess brightness and area indices will be briefly introduced. These disk-resolved indices will be contrasted with earlier implementations of indices, before a correlation analysis is performed, establishing the relation between disk-resolved and spectral activity indices.

%-------------------------------------------------------------------------------
%   Chromospheric Telescope
%-------------------------------------------------------------------------------

\subsection{Chromospheric Telescope}\label{SEC41}

In synoptic observing mode, ChroTel acquires one image per wavelength channel every three minutes. In good weather conditions, time-series of \cak\ filtergrams cover up to seven hours per day. The automated data-reduction pipeline includes corrections for dark- and flat-field frames, which yields Level~1.0 science-ready data in FITS format \citep{Bethge2011}. Corrections for image rotation and solar position angle still need to be applied. This study utilizes \cak\ filtergrams recorded between 2012~April and 2019~October (see Figure~1 in \citet{Diercke2019} for data coverage). The Median Filter-Gradient Similarity \citep[MFGS,][]{Deng2015, Denker2018} image quality metric was applied to the daily time-series to identify the best frame for each day. This image quality metric, among others, is implemented as part of the \textit{sTools} software library \citep{Kuckein2017}. The MFGS values for each filtergram are computed in a region of $600 \times 600$ pixels at disk center. The filtergrams with highest MFGS value is selected for further processing. 

Downsampling the 2048$\times$2048-pixel ChroTel images so that the solar radius corresponds to 512 pixels improves the SNR and yields an image scale of about 1.86\arcsec\ pixel$^{-1}$. In the next step, a two-dimensional limb-darkening function is computed following the recipe described in \citet{Denker1999}, which also includes an intensity normalization so that the quiet-Sun disk-center intensity corresponds to unity. Removal of the CLV results in a contrast-enhanced \cak\ image. 

Imaging of a large field-of-view (FOV) with Lyot filters often introduces variations of the background, which cannot be removed by flat-fielding and may even be time-dependent for a telescope with alt-azimuth mounts such as ChroTel. This problem was addressed by \citet{Shen2018} who demonstrated that field-dependent background variations can be fitted by Zernike polynomials \citep{Noll1976} so that the artifacts can be subsequently removed. Neglecting or overfitting the background variations significantly affects threshold-based activity indices. Choosing Noll's mode-ordering number $j = 21$ ($5^\mathrm{th}$ radial order) for fitting the ChroTel images ensures a good approximation of the background variations. The sensitivity of Zernike polynomials with higher values of $j$ is strongly biased towards the limb and does not match the spatial scale of the background variations. The background correction results in lower excess emission indices of about 8.0\%, whereby indices with higher thresholds are more strongly affected. In summary, the hitherto described procedures yield 892 contrast-enhanced, full-disk \cak\ filtergrams covering 22\% of Solar Cycle~24. As an example, two \cak\ filtergrams are displayed in the bottom panels of Figure~\ref{FIG02}{\ns} illustrating moderate activity and quiet-Sun conditions.

\citet{Johannesson1998} presented three indices based on contrast-enhanced \cak\ full-disk images obtained at BBSO. These data consist of video and digital records, which require different index definitions \citep{Johannesson1995}. The common criterion is an intensity threshold for feature selection, which is then used to compute the integral value of the excess emission. This yields an activity index accounting for the variation of the \cak\ irradiance. Continuing this effort, \citet{Naqvi2010} processed about 10 years of BBSO \cak\ filtergrams and computed the integral pixel brightness $D$, where the index definition followed that of \citet{Johannesson1998}. Furthermore, \citet{Naqvi2010} presented a new index, i.e., the area index $A$, which quantifies the fraction of the solar disk covered by features that made up the $D$-index. 

%-------------------------------------------------------------------------------
%   Figure 9 
%-------------------------------------------------------------------------------
\begin{figure}[t]
\includegraphics[width=\columnwidth]{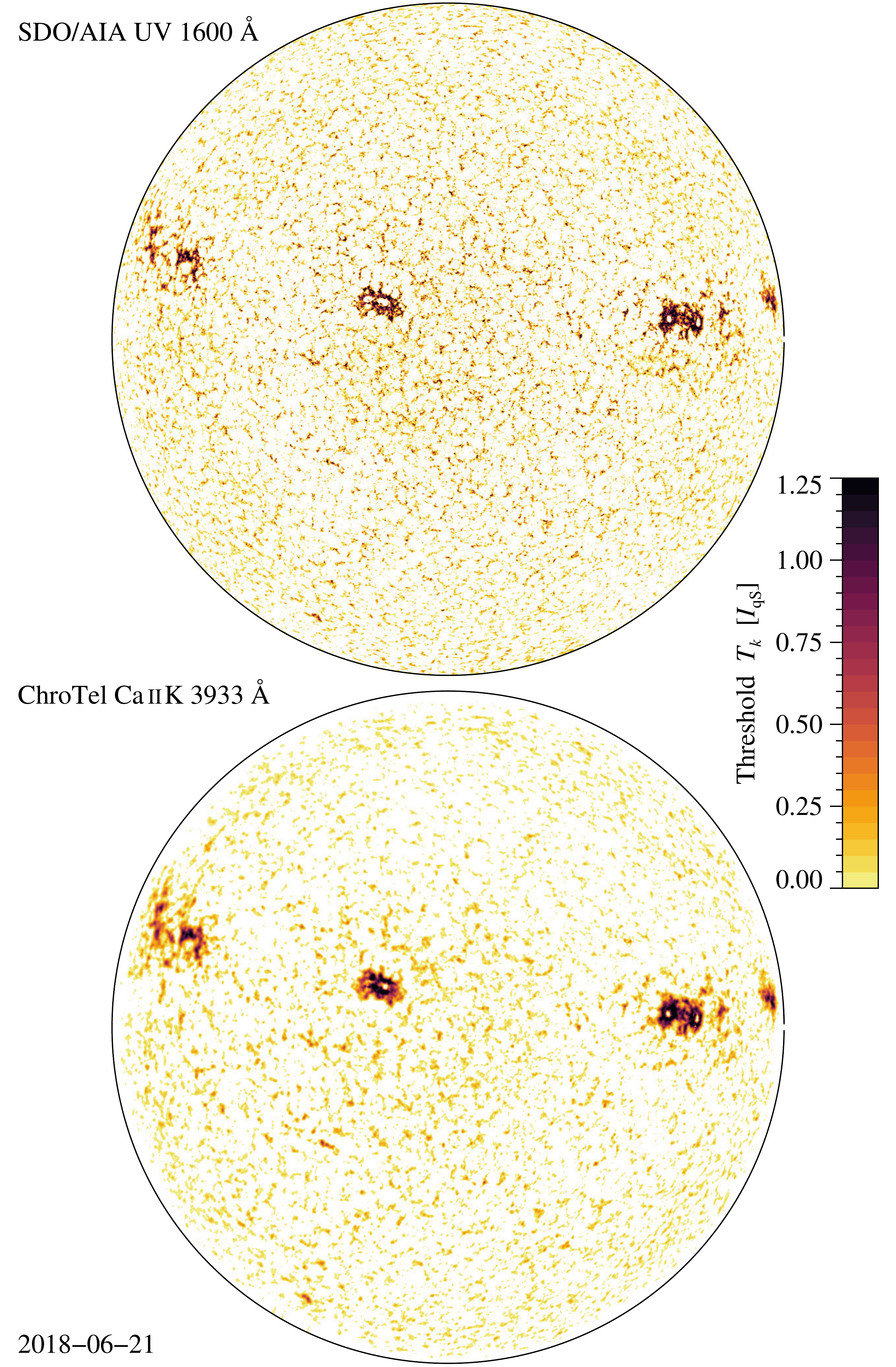}
\caption{Color-coded maps of the excess brightness thresholds $T_k$ for the 
    AIA UV 1600~\AA\ and the Chro\-Tel \cak\ 3933~\AA\ images on 2018 June~21 depicted in Figure~\ref{FIG02}{\ns}. }
\label{FIG08N}    
\end{figure}
%-------------------------------------------------------------------------------

Using enhanced-contrast \cak\ filtergrams, \citet{Naqvi2010} defined the excess brightness as
\begin{equation}
D_k = \frac{1}{1000} \sum_{ij} f_{ij}\ \mathrm{with}\left\{ 
    \begin{array}{ll}
    f_{ij} = R_{ij} - T_k & \ R_{ij} \geq T_k\\
    f_{ij} = 0            & \ R_{ij} < T_k
    \end{array}\right.\!\!,
\label{EQN4}
\end{equation}
where $R_{ij}$ is the intensity excess of a pixel and $T_k$ refers to an intensity threshold given as a fraction of the normalized quiet-Sun intensity \mbox{$I_\mathrm{qS} = 1$} at $\mu = 1$. The bottom panel of Figure~\ref{FIG08N}{\ns} is a visual representation of the thresholded excess brightness map. The intensity fraction is multiplied by 1000 for compact labeling of the excess brightness, i.e., $D_{100}$ corresponds to a threshold $T_{100} = 0.1$ or 10\% of the normalized disk-center quiet-Sun intensity $I_\mathrm{qS}$. The subscripts $i$ and $j$ represent the locations of pixels in Cartesian sun-center coordinates. The summation is carried out over all pixels contained within the solar disk. The division by 1000 casts the $D_k$-index values in a convenient data range. However, this has the drawback that the size of the full-disk images affects the amplitude of the $D_k$-indices. 

This issue can be addressed by using the empirical two-dimensional limb-darkening function. Summing the corresponding intensities of all pixels on the disk provides an estimate of the total brightness of the ``featureless'' Sun $L = \sum L_{ij}$. Therefore, the excess brightness in Equation~\ref{EQN4} can be expressed in terms of the total brightness
\begin{equation}
C_k = \frac{1}{L}\sum_{ij} f_{ij}.
\label{EQN5}
\end{equation}
The fraction $C_k$ can also be expressed as a percentage of the total brightness $L$. A similar approach can be used for area indices by summing all pixels for a certain threshold $T_k$ and dividing them by the number of all pixel contained with the solar disk $N$. Thus, $A^C_k$ refers to the fraction of the solar disk that is covered by features exceeding the brightness threshold $T_k$. The superscript $C$ is used for disambiguation in the later discussion of UV indices. This procedure assists in the physical interpretation of the $C_k$- and $A^C_k$-indices. A further qualitative improvement beyond the previous work \citep{Johannesson1995, Johannesson1998, Naqvi2010} is the removal of the field-dependent background variations (see Figure~\ref{FIG08N}{\ns} for validation) introduced by Lyot-type filters. Such artifacts often lead to a bias, especially near the limb, for low thresholds $T_k$ (cf.\ Figure~2 in \citet{Naqvi2010}). The remaining challenge for comparing disk-resolved \cak-indices is mainly related to the different passbands of the narrow-band \cak-filters, i.e., the narrower the filters, the higher the contrast of features such as plages \citep[cf.][]{Chatzistergos2020} and consequently the higher the excess brightness.

The BBSO \cak\ filtergrams, on which \citet{Naqvi2010} based their $D_k$-indices, were recorded with a 1.5~\AA\ (FWHM) Daystar filter. The  0.3~\AA\ (FWHM) ChroTel filter gathers light entirely from the chromosphere above the temperature minimum. Consequently, a ten times higher $T_k$ corresponds roughly to the indices described in previous works \citep{Johannesson1995, Johannesson1998, Naqvi2010}. The thresholds are mapped onto the full-disk images in Figure~\ref{FIG08N}{\ns} and range between 5\% and 125\% of the normalized quiet-Sun intensity $I_\mathrm{qS}$ at disk center. According to the above nomenclature, the $C_{1000}$-index computed with a threshold of $T_{1000} = 1$ or 100\% of $I_\mathrm{qS}$ has a corresponding area-fraction index $A^C_{1000}$. The $C_{1000}$-index includes the excess emission of only the brightest part of active region plages. 

The excess brightness index $C_{0}$ includes the entire chromospheric network and all plage regions without any distinction between these features. Increasing the threshold gradually reduces contributions from the dimmer quiet network, leaving only islands of bright magnetic features. Thus, the $C_{250}$-index includes plage regions with different sizes and the enhanced network, which appears as an ``orange peel'' pattern of the supergranular boundaries in Figure~\ref{FIG08N}{\ns}. The next index $C_{500}$ contains only the brightest elements of the enhanced network, whereas the $C_{750}$-index consists mainly of plages. These thresholds are only indicative, i.e., without information about magnetic field and local dynamics, it is impossible to properly differentiate between active features. 

%-------------------------------------------------------------------------------
%   Figure 10 
%-------------------------------------------------------------------------------
\begin{figure}[t]
\includegraphics[width=\columnwidth]{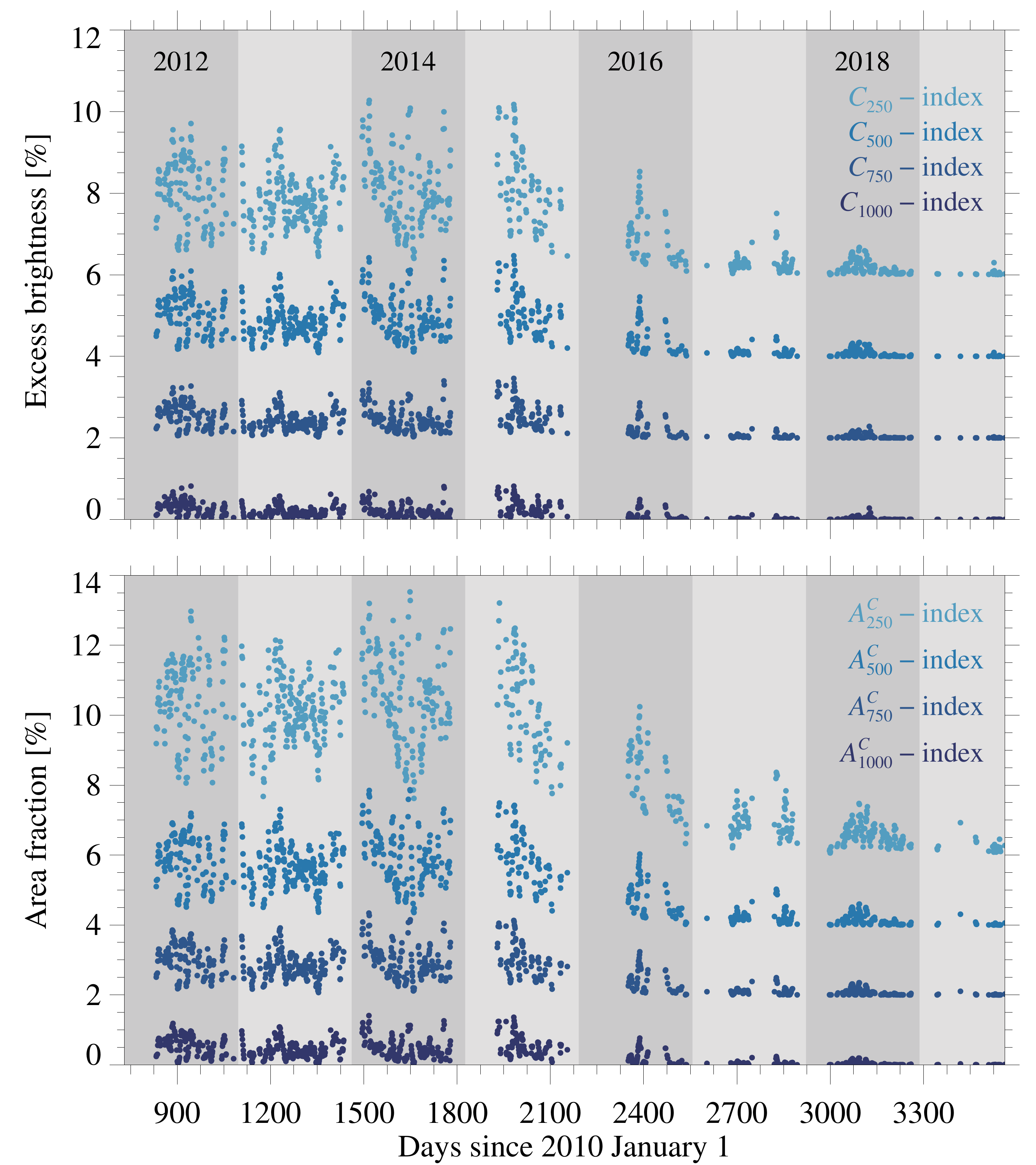}
\caption{Time-series of the excess brightness indices $C_k$ and their 
    corresponding area indices $A^C_k$ based on ChroTel \cak\ full-disk images. The alternating dark and light gray stripes mark successive years since 2010. Each index pair $C_k$ and $A^C_k$ is plotted in the same shade of blue. The legend in the upper-left corner of the panels provides the key to the intensity thresholds of 25\%, 50\%, 75\%, and 100\%. The indices are vertically separated by a constant value for better display. The gaps in the datasets correspond to days without observations, and some days with mediocre data quality were removed as well. }
\label{FIG10}    
\end{figure}
%-------------------------------------------------------------------------------

The aforementioned four subsets of the excess brightness time-series $C_k$ are presented in the top panel of Figure~\ref{FIG10}{\ns} along with the corresponding $A^C_k$-indices. The indices are expressed as a percentage of total brightness $L$ of the featureless Sun and of the solar disk area, respectively. The $C_{250}$-index show the largest amplitude since it include most of the magnetic features seen in the \cak\ line. This is also the only index displayed in Figure~\ref{FIG10}{\ns}, which does not reach zero values. The amplitude of each following $C_k$-index reduces monotonically, where $C_{250}$-, $C_{750}$-, and $C_{1000}$-indices reach values below 0.1\% also during periods of high solar activity. All four indices show peaks corresponding to the maximum of Solar Cycle~24 around 2014~April and strong secondary peaks in 2014~July and October. The ChroTel excess brightness indices represent well the general secular trend of Solar Cycle~24 \citep[e.g.,][]{Chatzistergos2020}. However, the exact time of the peaks is difficult to determine due to gaps in the time-series.

The appearance of the $A^C_k$-indices resembles its companion $C_k$-indices, however with larger amplitudes. Moreover, the maximum of the $A^C_{250}$-index occurs in 2014~July accompanied by two secondary peaks in 2014~March and October. The excess brightness and area fraction indices show significant rotational modulations. The strong peaks around solar maximum can be attributed to large, long-living, magnetically complex active regions passing across the solar disk. Growth and decay of these recurring active regions leave their marks as secondary maxima.

%-------------------------------------------------------------------------------
%   Atmospheric Imaging Assembly
%-------------------------------------------------------------------------------

\subsection{Atmospheric Imaging Assembly}\label{SEC42}

The next set of excess brightness and area indices is based on long-integration UV images. Before computing the time-averaged UV images, each single-exposure image is corrected for differential rotation, with the 12:00~UT frame as a reference. Furthermore, the long-integration images are normalized such that the disk-center intensity is unity, and in the next step, the two-dimensional CLV function is subtracted.

Each of the hereto outlined steps are executed analogous to the processing of the ChroTel \cak\ filtergrams described in Section~\ref{SEC41}. The AIA observations are not compromised by Earth's turbulent atmosphere and instrument jitter. Thus, the UV 1600~\AA\ full-disk images have intrinsically a better contrast than the ground-based ChroTel observations. In addition, the filter transmission is much more uniform compared to the ChroTel Lyot filter, so that the excess emission is on average only about 1.6\% lower compared to images without background correction. According to \citet{Simoes2019}, the plage emission in the AIA 1600~\AA\ channel is dominated by photospheric continuum emission. The bright features trace the underlying structure of chromospheric \cahk\ plages and network, and to lesser extent the internetwork grains \citep{Tritschler2007, Krijger2001}. 

Deriving the UV indices follows the same logic as presented in \mbox{Section~\ref{SEC41}}. First, the excess brightness is computed according to \mbox{Equation~\ref{EQN4}} employing thresholds $T_k$ between 0\% and 125\% of the normalized quiet-Sun intensity $I_{qS}$. The AIA long-integration, contrast-enhanced UV full-disk images adapt to the same brightness thresholds as used for the $C_k$-index. A map of the solar disk with brightness thresholds mapped onto it is displayed in the top panel of Figure~\ref{FIG08N}{\ns}. 

Finally, according to Equation~\ref{EQN5}, the UV excess brightness $U_k$-index is expressed as a fraction of the total brightness of the ``featureless'' Sun $L_U$ with an associated area $A^{U}_k$-index. Outliers in the time-series are identified as values above the $3\sigma$-level computed for the entire detrended dataset. In addition, only 37 days are missing for the period since the start of AIA observations until 2019 November~26. The missing data points are replaced by median values to avoid gaps in the otherwise continuous index time-series. 

%-------------------------------------------------------------------------------
%   Figure 11
%-------------------------------------------------------------------------------
\begin{figure}[t]
\includegraphics[width=\columnwidth]{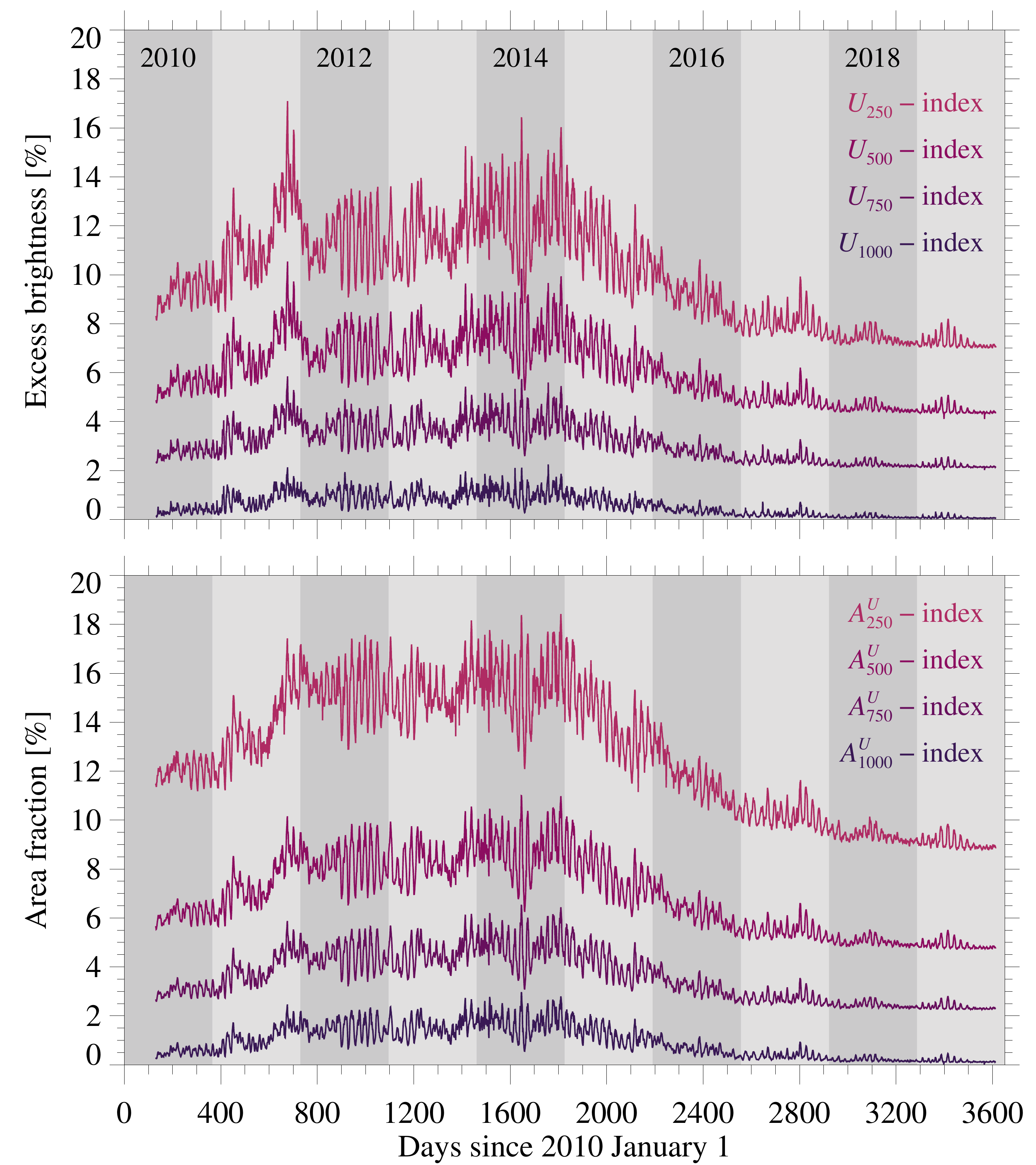}
\caption{Time-series of the excess brightness indices $U_k$ and their 
    corresponding area indices $A^U_k$ based on AIA 1600~\AA\ full-disk images. The alternating dark and light gray stripes mark successive years since 2010. The index pairs are plotted in the same shade of purple. The legend in the upper-left corner of the panels provides the key to the intensity thresholds of 25\%, 50\%, 75\%, and 100\%. The indices are vertically separated by a 2\% for better display. }
\label{FIG11}    
\end{figure}
%-------------------------------------------------------------------------------

The AIA observations cover most of Solar Cycle~24, through a gradual rise of activity from 2010 to 2014, when the cycle reaches its maximum, followed by a steeper ascending phase towards the minimum in 2019 November. This development is well represented in the four subsets of the $U_k$-indices and their $A^{U}_k$-index companions displayed in Figure~\ref{FIG11}{\ns}. Similar to Figure~\ref{FIG10}{\ns}, these plots correspond to the brightness thresholds $T_k$ equal to 25\%, 50\%, 75\%, and 100\% of the normalized disk-center intensity $I_{qS}$. 

%-------------------------------------------------------------------------------
%   Figure 12 
%-------------------------------------------------------------------------------
\begin{figure}[t]
\includegraphics[width=\columnwidth]{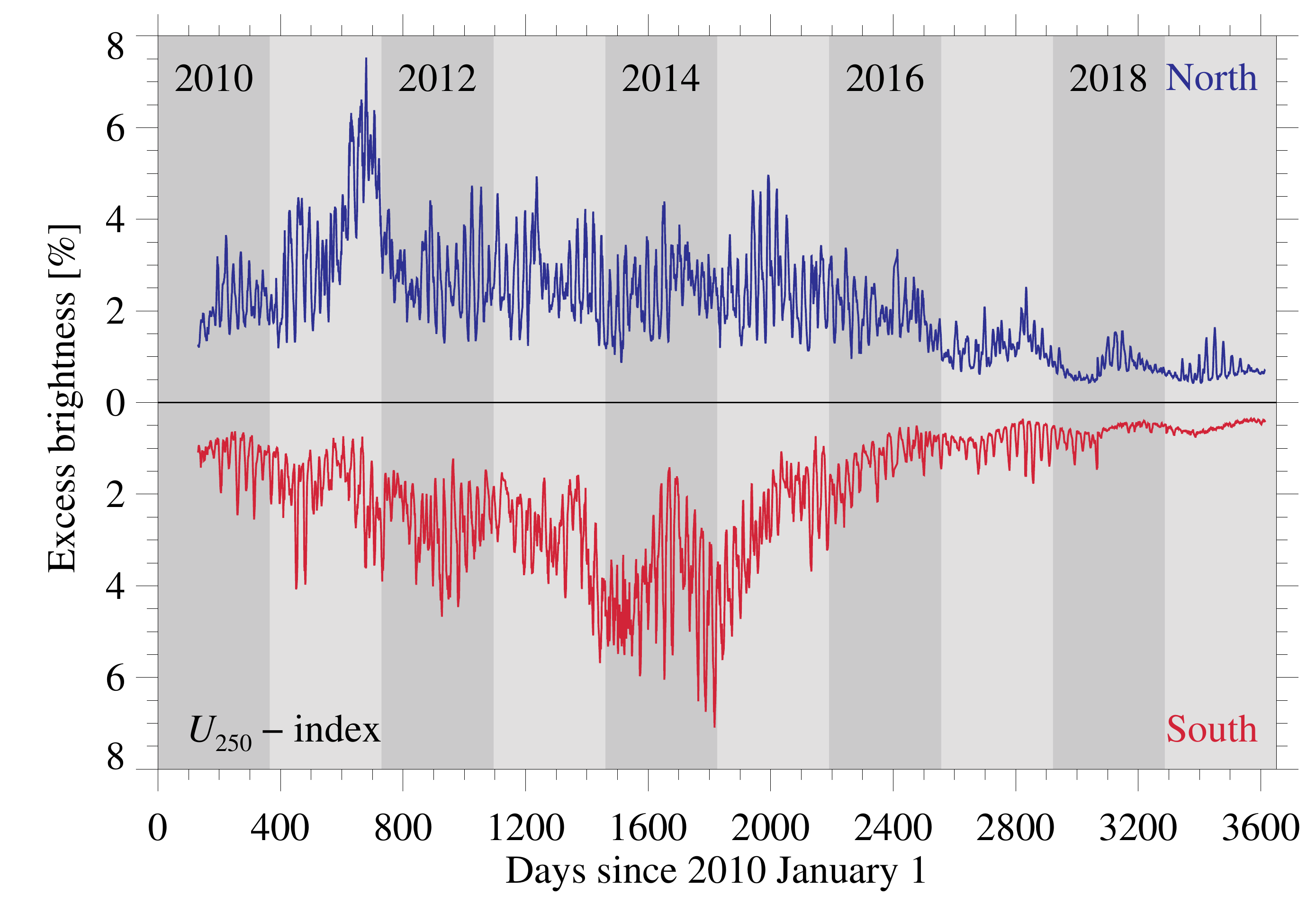}
\caption{Time-series of the excess brightness $U_{250}$-index computed 
    separately for the northern and southern hemispheres based on AIA UV 1600~\AA\ long-integration images with a 25\% threshold. The alternating dark and light gray stripes mark successive years since 2010. The southern hemispheric $U_{250}$-index is plotted upside-down for convenience. }
\label{FIG12}    
\end{figure}
%-------------------------------------------------------------------------------

The $U_{250}$-index, which combines the excess brightness of network and plages alike, varies between 1\% and 10\%. Reducing the contribution of the network, the $U_{500}$-index varies between 0.1\% and 6.3\% and the $U_{750}$-index between 0.02\% and 3.74\%. Finally, the $U_{1000}$-index, which contains only the brightest elements of plages and active regions, varies between 0.003\% and 2.25\%. All four $U_k$-indices show the same secular trend, but with reduced amplitude when increasing the threshold. The similarity of the $U_{250}$- and $U_{1000}$-indices suggests that the main drivers of the overall UV variability is the excess UV emission of large complex active regions, whereas the network is responsible for the basal emission. The same conclusion holds true for the chromosphere, as becomes evident from a comparison of Figures~\ref{FIG10}{\ns} and~\ref{FIG11}{\ns}. 

The high level of activity that is present in all $U_k$-indices, culminating in late 2011, encompasses a period of about seven months, where several active regions evolve \citep{Song2018}, which are surrounded by bright plages and enhanced network. The $A_k$-index time-series shows the larger coverage by active regions as well but at a reduced level, which stays below the level for the secondary maximum of solar activity in 2014. The very high $U_k$-values in late 2011 are an indication that the intrinsic excess brightness of active region plages is enhanced during solar maximum conditions, which cannot be explained by a larger coverage alone. 

%-------------------------------------------------------------------------------
%   Figure 13
%-------------------------------------------------------------------------------
\begin{figure}[t]
\includegraphics[width=\columnwidth]{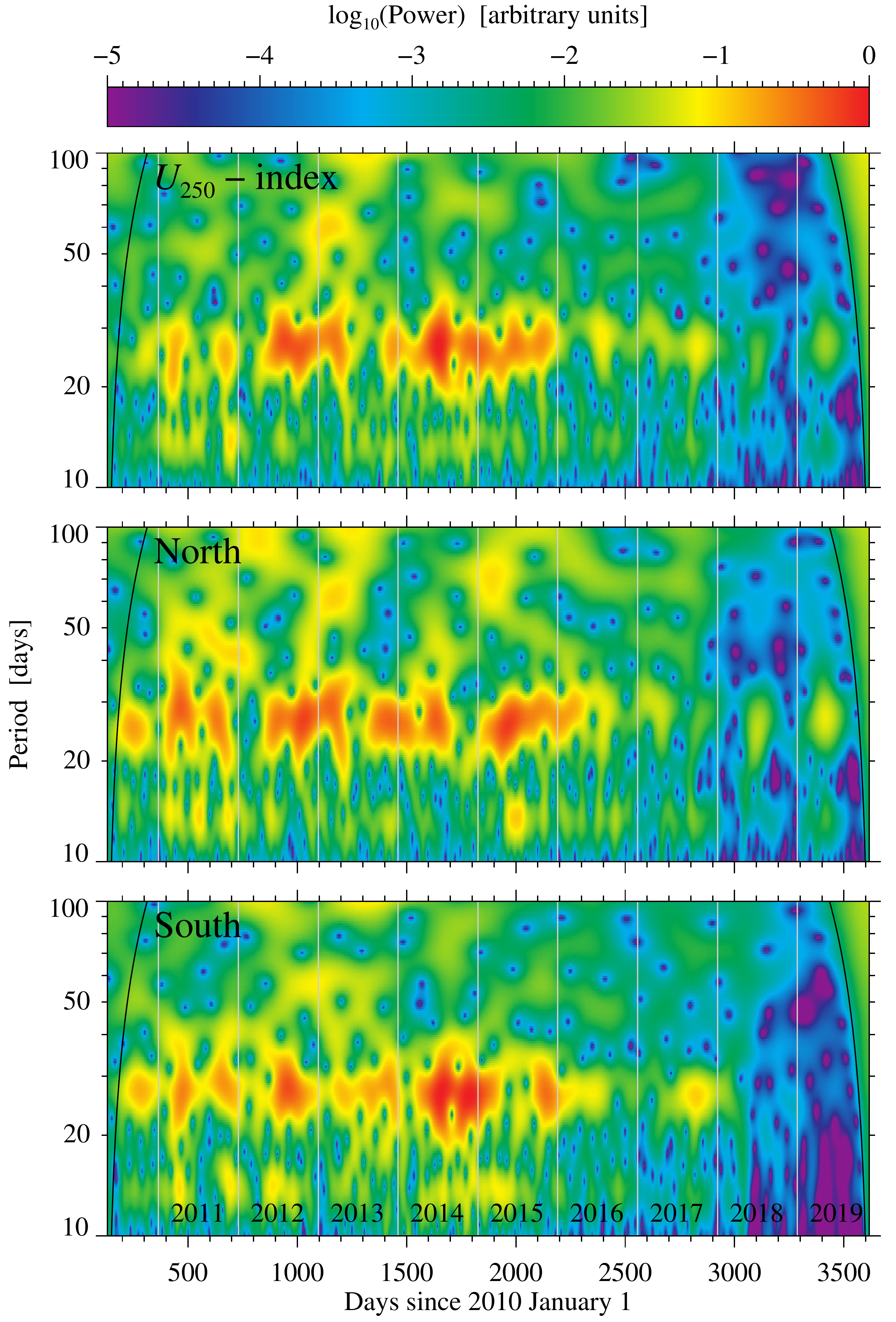}
\caption{Power spectrum based on the Morlet wavelet transform of the global
    $U_{250}$-index and its two hemispheric components. The two curved lines at the sides represent the cone-of-influence, i.e., outside these bounds, edge effects impinge upon the power spectrum. The rainbow-colored logarithmic scale of the power spectral density covers five decades.}
\label{FIG13}    
\end{figure}
%-------------------------------------------------------------------------------

The $U_{250}$-index computed separately for each hemisphere is presented in Figure~\ref{FIG12}{\ns}. The southern hemisphere index is plotted upside-down to visualize and compare the temporal variations of $U_{250}$ index in both hemispheres. The asymmetry of solar cycle variations is striking. Both hemispheres evolve independently, and the presence of large active regions have a considerable impact. The northern hemisphere shows a first maximum in late 2011, whereas a double-peaked maximum appears in 2014 in the southern hemisphere. Solar activity is high in the southern hemisphere for the entire year 2014. This trend reverses in early 2015, but at lower activity levels. Similar trends are observed in the SSN time-series and also the hemispheric sunspot number (HSN) exhibits asymmetries \citep{Singh2021a, Veronig2021}. However, a certain lag between the SSN and the excess brightness time-series is expected, e.g., due to the differences in lifetime and decay rates of bipolar active regions and plages \citep{Foukal1998}.

Knowledge about hemispheric asymmetries is essential in dynamo theory. In particular, low-latitude active regions have the potential to exchange flux across the equatorial boundary, affecting flux transport to the poles by meridional flows. The rotational signatures are well defined in both hemispheric time-series. However, starting in 2018, the rotational modulation of the $U_{250}$-index almost completely vanishes from the southern hemisphere while still being present at a low level in the northern hemisphere \citep[cf.][]{Sun2015}. Thus, the southern hemisphere may represent in this period the lowest activity state of the Sun observed so far. 

\citet{Denker2019} presented a magnetic index based on BaSAMs employing HMI line-of-sight (LOS) magnetograms. A comparison with Figure~10 in \citet{Denker2019} shows that the excess brightness $U_{250}$-index follows closely the magnetic BaSAM-index rather than the daily sunspot number. For example, both indices show the two maxima in 2012 and 2014, where the latter one is double-peaked, which are unique characteristics of Solar Cycle~24. Furthermore, the $U_{250}$-index maintains a basal level, much like the magnetic BaSAM-index.  Thus, the magnetic and excess brightness indices represent ``two sides of the same coin'', i.e., the variation of the ubiquitous magnetic field of the Sun plus the contributions by (decaying) active regions. 

The wavelet power spectrum displayed in Figure~\ref{FIG13}{\ns} also displays the signature of the aforementioned North-South asymmetry. The top panel shows the power spectrum of the global $U_{250}$-index, whereas the middle and bottom panels correspond to the northern and southern hemispheric $U_{250}$-indices, respectively, as displayed in Figure~\ref{FIG12}{\ns}. Both hemispheric power spectra are normalized by the same factor, i.e., the average of the maximum power in both hemispheres, ensuring that the power spectrum maps can be directly compared. The rainbow colors are indicative of the power spectral density as a function of period and observing date. Bands of high power (yellow and red colors) occur a periods close to the Carrington rotation period. 
%-------------------------------------------------------------------------------
%   Figure 14 
%-------------------------------------------------------------------------------
\begin{figure*}[t]
\includegraphics[width=\textwidth]{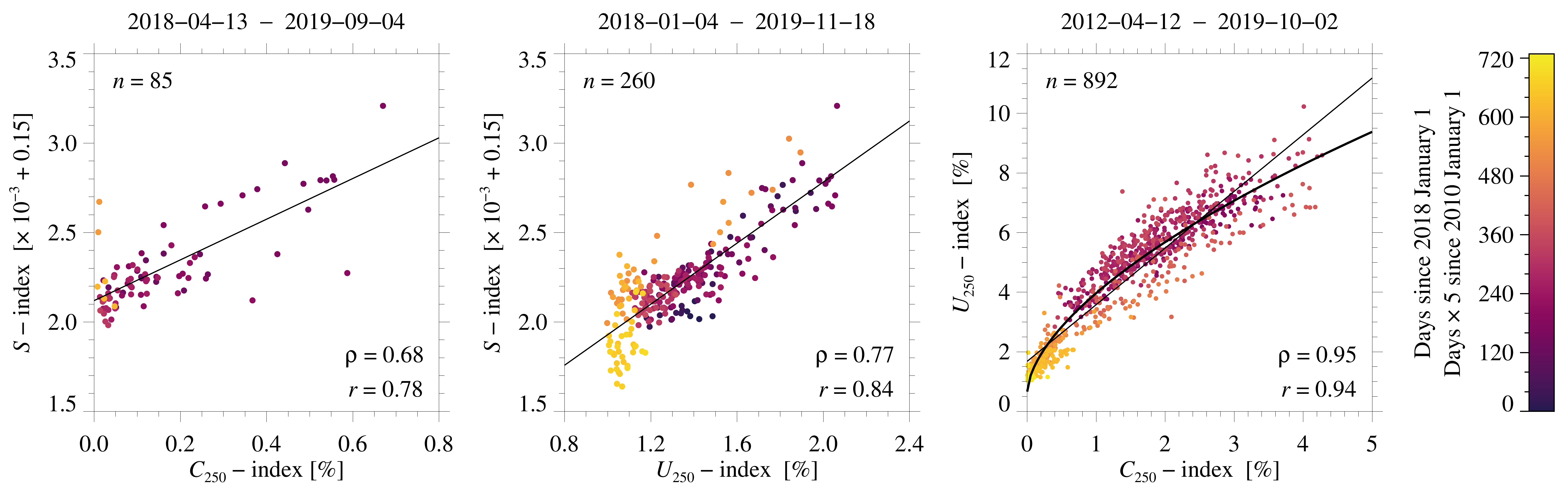}
\caption{Scatter plots of the $S$-, $C_{250}$-, and $U_{250}$-indices. The 
    diagonal straight lines represent linear fits. In addition, a geometric regression model was applied to the $U_{250}$-index (\textit{thick curve}). Pearson's linear and Spearman's rank-order correlation coefficients $r$ and $\rho$ are given in the lower-right corner of each panel, respectively. The number of common data values $n$ differs in each plot, and their chronological order is color-coded. Dates of the first and last data points are given above the individual plot panels.}
\label{FIG14}    
\end{figure*}
%-------------------------------------------------------------------------------
The northern $U_{250}$-index has a steady rotational modulation pattern, separated into compact groups associated with large active regions rotating on and off the solar disk. Therefore, the wavelet power spectrum exhibits several strong cores related to the Carrington rotation period. The main rotational signal in the southern hemisphere also appears around the Carrington rotation period. However, the power amplitude is significantly lower, except the two-peaked maximum of solar activity in 2014. The width of the bands with enhanced power is about 10~days, though broadening by excursions to lower periods exists. Power at longer periods is mainly observed in the northern hemisphere but at a lower level and with a more diffuse appearance. Short-period power occurs at half the Carrington rotation period and likely corresponds to the disk passage of isolated and short-lived active regions. Hemispheric asymmetries cannot be deduced from the global power spectrum, which accentuates the maxima in 2012 and 2014 for the northern and southern hemispheres, respectively. The drop of power by about five orders of magnitude is a distinguishing feature for the southern hemisphere during the solar minimum period in 2018 and 2019. The rotational modulation pattern is completely absent, while still being present at a  much reduced level in the northern hemisphere.

%-------------------------------------------------------------------------------
%  Spectral vs. disk-resolved indices
%-------------------------------------------------------------------------------

\subsection{Spectral vs.\ disk-resolved indices}\label{SEC43}

Two major types of activity indices are discussed in this work. The $S_\mathrm{PEPSI}$-index belongs to the first type, i.e., a Sun-as-a-star spectral index, whereas the $C_k$- and $U_k$-indices constitute the second type, i.e., an index that amalgamates information from disk-resolved data. Combined, these indices facilitate a robust multi-wavelength analysis of the solar activity cycle. While plages dominate the \cahk\ and UV 1600~\AA\ continuum emissions during high solar activity, the magnetic network sustains the basal emission during low activity periods. Consequently, these indices are proxies of the magnetic field strength, and the different thresholds isolate different types of magnetic field topology. Thus, they provide clues about the coupling between photosphere and chromosphere and the resulting energy transport. 

The $S_\mathrm{PEPSI}$-index benefits from its heritage as a Sun-as-a-star index, combining the line-core \cahk\ emission. However, the line-core passbands, as defined in \citet{Vaughan1978}, include certain contributions from the photospheric line wings. Therefore, the $S$-index is not entirely a chromospheric activity index. According to \cahk\ formation theory \citep{Bjorgen2018}, the $S$-index describes the combined magnetically driven emission variability of the upper photosphere, around the temperature minimum, up to about 2000~km above the solar surface. 

The narrow \cak\ passband isolates light originating only from above the temperature minimum. Therefore, this excess brightness index can be considered purely chromospheric. Its UV counterpart gathers light originating from the temperature minimum region \citep{Vernazza1976, Vernazza1981, Fontenla1993}. 

The $S_\mathrm{PEPSI}$-, $C_{250}$-, and $U_{250}$-indices for the period between 2018 January~1 and 2019 December~31 are displayed in Figure~\ref{FIG03}{\ns}. Beyond different technical and methodological implementations, similarities and differences in the three indices are driven by the physical properties of the respective radiation formation heights. The $U_{250}$-index presents the most complete record. Therefore, it is used as the benchmark time-series. 

Just considering the years 2018 and 2019, i.e., towards the end of Solar Cycle 24, the highest activity level in the \cak\ disk-resolved index occurs in 2018~June. Although the $C_{250}$-index does not reach exactly zero, it closely approaches it during periods of low activity. On the other hand, the $S_\mathrm{PEPSI}$- and $U_{250}$-indices have a well-defined basal level, which stay above zero during low solar activity. Similarities between the $S_\mathrm{PEPSI}$- and $U_{250}$-indices may arise from the photospheric contamination of the $S$-index and a more closely matching formation height. 

The relationship among the three indices is illustrated in three scatter plots displayed in Figure~\ref{FIG14}{\ns}. The first two panels compare index values in the period displayed in Figure~\ref{FIG03}{\ns}, which is defined by the length of the \mbox{PEPSI/SDI} time-series. The first panel shows the correlation between the two \cak\ indices. A lower rank-order correlation coefficient $\rho$ indicates a less monotonic relationship among the indices. In this case, it points to a larger scatter so that a simple linear model is sufficient for the fit. In addition, considering the small sample of 85 common data points for the comparison of \mbox{PEPSI/SDI} and ChroTel data, a linear fit seems appropriate and demonstrates a good correlation of these datasets. The $S_\mathrm{PEPSI}$- and $U_{250}$-indices show a similar trend (middle panel) with better statistics (three times more data points) but with higher correlations. The orange and yellow data points show a larger scatter in the S-index at times of extremely low activity in 2019. This is also evident in the first panel, but only for a few data points because of the small temporal overlap of the two indices in 2019. In particular, $S$-index values are below the trend line in October and November~2019, whereas the values between June and September~2019 are on the same scale as the 2018 $S_\mathrm{PEPSI}$-index. This may indicate that active region plages and the low associated rotational modulation, seen in the $U_{250}$-index for this period, become inconsequential for the $S$-index.

The third panel compares a much larger sample ($n = 892$) of the overlapping $C_{250}$- and $U_{250}$-indices. The linear and rank-order correlation coefficients approach unity. A linear regression delivers already good results. However, a geometric regression model $y = a_0 + a_1x^{a_2}$ adapts better to data points at higher levels of the solar activity. The fitted exponent $a_2 = 0.6 \approx 0.5$ implies a quadratic relationship between the $C_{250}$- and $U_{250}$-indices. A comparison of Figures~\ref{FIG10}{\ns} and~\ref{FIG11}{\ns} shows that the evolution with time is similar with respect to the secular trend and the rotational modulation. Therefore, the quadratic relationship indicates topological differences in the emitting features. These differences become smaller during low activity, when the $C_{250}$- and $U_{250}$-indices mainly describe the quiet chromospheric network, proving that short- vs.\ long-term relations are not necessarily identical \citep{Pevtsov2014}. Increasing activity leads to a higher topological complexity of the magnetic field and canopy effects, which are more prominent in the \cak\ filtergrams, become more important.

%===============================================================================
%   CONCLUSIONS 
%===============================================================================

\section{Conclusions}\label{SEC5}

The Mt.\ Wilson HK-monitoring program started a multi-decade effort, which continuous to this day, of solar and stellar \cahk\ spectroscopy as a means to identify and characterize magnetic activity cycles in main-sequence stars, e.g., via the $S$-index. The analysis of the thus compiled data contributed significantly to our current knowledge of stellar activity from quiet over periodic to highly active stars and to a better understanding of stellar atmospheres. Recognizing signs of activity in the stellar spectrum is crucial when applying the radial velocities method in the search of exoplanets \citep{Astudillo_Defru2017, Dumusque2018}.

The HEROES $S$-index \citep{Schroder2012} and the $S_\mathrm{PEPSI}$-index (see Section~\ref{SEC31}) show similar levels of activity, albeit focusing on two different minima, i.e., at the beginning and end of Solar Cycle~24. In a follow-up study, \citet{Schroder2017} pointed to a decline of the HEROES $S$-index starting in 2015 and reaching values below 0.160 already in 2016. The \citet{Maldonado2019} HARPS $S$-index time-series (also discussed in Section~\ref{SEC31}) confirms the \citet{Schroder2017} results for the time when both time-series overlap. Based on the time-series ending in 2016, \citet{Schroder2017} argue that Cycle~24 is a weak activity cycle with small amplitude. This is corroborated by the average $S_\mathrm{PEPSI}$-index of about 0.152 in 2018 and 2019, which reflect some of the lowest $S$-index values reported so far. The declining trend of the $S_\mathrm{PEPSI}$-index continues towards solar minimum at the end of 2019. The composite $S$-index time-series presented in \citet{Egeland2017} indicates that $S$-index values below 0.160 are rare during solar minima. The deepest minimum occurred between Solar Cycles~22 and~23. However, comparison with previous activity cycles must be made with caution because the SNR and spectral resolution of the spectra significantly affect the $S$-index (see discussion in Section~\ref{SEC33}).

The Sun's special status in chromospheric activity studies arises from the fact that the spectral ``fingerprint'' left in Sun-as-a-star observations can be matched to activity features on the solar disk. This provides guidance for interpreting stellar spectra exhibiting signatures of chromospheric activity. The $C_{250}$-index, which includes all bright features of the active Sun, indicates an almost complete absence of plages and enhanced network elements -- in particular in the southern hemisphere during 2019. Thus, the $S_\mathrm{PEPSI}$-index at that time represents the basal chromospheric activity during a deep solar minimum.

Regular PEPSI/SDI observation with high-spectral resolution produced a daily though irregular record, which, combined with elaborate data-processing techniques and analysis, yields a versatile dataset with broad spectral coverage including the strong \cahk\ absorption lines. Other strong chromospheric absorption lines such as the Balmer series, the infrared Ca\,{\sc ii} triplet or the Na\,D doublet were not yet explored and promise more information on solar cycle variations of the chromosphere. Thus, this work is only the initial pass of an ongoing effort to characterize solar activity via its PEPSI/SDI spectra. Furthermore, the existence of synoptic full-disk observations from ground- and space-based instrumentation presents a unique, i.e., in the stellar context, opportunity to study the surface features, which are responsible for the spectral signatures that modulate the $S$-index. The $S_\mathrm{PEPSI}$-index is computed for the declining phase of the Solar Cycle~24. Various data analysis techniques were employed and demonstrated that a strong relationship exists between the $S$-index and the \cak\ and UV 1600~\AA\ continuum excess brightness. 

The analysis of the disk-resolved indices during the recent low-activity period of Solar Cycle~24 confirms the role of the chromospheric network as the major contributor to the basal \cahk\ emission. It is the manifestation of the perpetual chromospheric magnetic field. Considering age and rotation of Sun-like stars with comparable activity levels \citep{Baliunas1995, Hall2007}, basal \cahk\ emission is always present, even for stars with a flat activity level, which is also recognized in theoretical descriptions  \citep[e.g.,][]{Schrijver1989, Schroder2012}. However, the combination of disk-integrated spectra and full-disk observations provided physical parameters that made it possible to address scenarios with different contributors to chromospheric activity. 

Studying long-term solar activity based on disk-resolved observations from various instruments is a complicated and nontrivial task \citep{Chatzistergos2020}. For example, to relate the ChroTel $C_k$-index to the BBSO $D_k$-index \citep{Naqvi2010} faces the challenge that observations do not overlap. Fortuitously, the synoptic SDO observations comprise a wealth of full-disk filtergrams (visible, UV, and EUV), magnetograms, and dopplergrams. In particular, the AIA UV 1600~\AA\ proved to be very helpful for comparing the $S_\mathrm{PEPSI}$-index and the ChroTel $C_k$- and $A_k$-indices. More generally, the AIA observations can serve as a straightforward benchmark, when comparing solar activity indices for most of Solar Cycle~24 and the just started Solar Cycle~25.

%===============================================================================
%    ACKNOWLEDGEMENTS
%===============================================================================

\section{Acknowledgments}

PEPSI and SDI were made possible by longtime support of the German Federal Ministry (BMBF) for the collaborative research (Verbundforschung) projects 05AL2BA1/3 and 05A08BAC. ChroTel is operated by the Leibniz Institute for Solar Physics (KIS) in Freiburg, Germany, at the Spanish Observatorio del Teide on Tenerife (Spain). The ChroTel filtergraph was developed by KIS in cooperation with the High Altitude Observatory (HAO) in Boulder, Colorado. SDO AIA data are provided by the Joint Science Operations Center -- Science Data Processing. This research has made use of NASA's Astrophysics Data System (ADS). This study was supported by grants DE~787/5-1 of the Deutsche Forschungsgemeinschaft (DFG). In addition, the support by the European Commission's Horizon 2020 Program under grant agreements 824064 (ESCAPE -- European Science Cluster of Astronomy \& Particle physics ESFRI research infrastructures) and 824135 (SOLARNET -- Integrating High Resolution Solar Physics) is highly appreciated. ED is grateful for the generous financial support from German Academic Exchange Service (DAAD) in the form of a doctoral scholarship. JP's research internship in Germany was made possible by the Research Internships in Science and Engineering (RISE) program of the German Academic Exchange Service (DAAD). MV acknowledges support by grant VE~1112/1-1 of the DFG. We would like to thank the referee for carefully reading our manuscript and for constructive comments, which substantially improved the contents of the article.

%\clearpage\newpage

%===============================================================================
%    BIBLIOGRAPHY
%===============================================================================

\end{document}